\documentclass[natbib]{aa} 

\usepackage{graphicx} 
\usepackage{txfonts} 
\usepackage{amsmath} 
\usepackage[table, dvipsnames]{xcolor}
\definecolor{lightgray}{gray}{0.9}
\definecolor{heavygray}{gray}{0.5}
\usepackage[colorlinks=True, allcolors=blue]{hyperref}
\usepackage{lscape}
\begin{document} 
\title{Stellar population astrophysics (SPA) with the TNG\thanks{Based on observations made with the Italian Telescopio Nazionale Galileo (TNG) 
operated on the island of La Palma by the Fundación Galileo Galilei of the INAF (Istituto Nazionale di Astrofisica) at the Spanish Observatorio 
del Roque de los Muchachos of the Instituto de Astrofisica de Canarias.
This study is part of the 
Large Program titled {\it SPA - Stellar Population Astrophysics:  the detailed, age-resolved
chemistry of the Milky Way disk} (PI: L. Origlia), granted observing time with HARPS-N and GIANO-B echelle spectrographs
at the TNG.}}
   \subtitle{The chemical content of the red supergiant population in the Perseus complex}

   \author{C. Fanelli \inst{1,2}
          \and
          L. Origlia\inst{2}
          \and
          E. Oliva\inst{3}
          \and
          E. Dalessandro\inst{2}
          \and
          A. Mucciarelli\inst{1,2}
          \and
          N. Sanna\inst{3}
          }

   \institute{Dipartimento di Fisica e Astronomia, Università degli Studi di Bologna, via Piero Gobetti 93/2, 40129, Bologna, Italy,\\
	\email{cristiano.fanelli3@unibo.it}
   \and
    INAF-Osservatorio di Astrofisica e Scienza dello Spazio, via Piero Gobetti 93/3, 40129, Bologna, Italy
    \and 
    INAF-Osservatorio Astrofisico di Arcetri, Largo Enrico Fermi 5, 50125, Firenze, Italy\\
             }
   \date{}
%
\abstract
{The Perseus complex in the outer disk of the Galaxy hosts a number of clusters and associations of young stars. Gaia is providing a detailed characterization of their kinematic structure and evolutionary properties.}
{Within the SPA Large Programme at the TNG, we secured HARPS-N and GIANO-B high-resolution optical and near-infrared (NIR) spectra of the young red supergiants (RSG) stars in the Perseus complex, in order to obtain accurate radial velocities, stellar parameters and detailed chemical abundances.} 
{We used spectral synthesis to best-fit hundreds of atomic and molecular lines in the spectra of the observed 27 RSGs. We obtained accurate estimates of the stellar temperature, gravity, micro and macro turbulence velocities and chemical abundances for 25 different elements. We also measured the $^{12}C/^{13}C$ abundance ratio.}
{Our combined optical and NIR chemical study provides homogeneous half-solar iron with a small dispersion, about solar-scaled abundance ratios for the iron-peak, alpha and other light elements and a small enhancement of Na, K and neutron-capture elements, consistent with the thin disk chemistry traced by older stellar populations  at a similar Galactocentric distance of about 10 kpc. 
We inferred enhancement of N, depletion of C and of the $^{12}$C/$^{13}$C isotopic abundance ratio, consistent with mixing processes in the stellar interiors during the RSG evolution.}
{}
\keywords{Techniques: spectroscopic - stars: abundances - stars: late-type - stars : supergiants}
\maketitle
\section{Introduction}\label{Intro}
The full characterization of the young stellar populations in recently formed star clusters and associations within the Galaxy is of paramount importance to answer a number of open questions concerning the formation and early evolution of star clusters and associations as well as the recent star formation and chemical enrichment history of the Milky Way disc.

The Gaia mission of the European Space Agency \citep{gaia_mission} is opening a new perspective in identifying and kinematically characterizing sub-structures in the inner and outer disc of the Galaxy. Follow-up spectroscopic observations at medium-high resolution provide complementary information on the detailed chemistry and line-of-sight radial velocities (RVs) of those regions.

The Perseus complex at a Galactocentric distance of about 10~kpc hosts a number of young star clusters and associations, the densest structures being h,$\chi$ Per, NGC~457, NGC~654 and NGC~633, distributed within a projected area of approximately 10 degrees on sky.
This region offers a unique opportunity to study the recent star formation and cluster assembling processes in the outer disc. 

Recently \citet{deburgos20} provided a comprehensive kinematic study of the young Perseus OB-1 association, by combining Gaia DR2 and high resolution optical spectroscopy, finding that most of the sampled blue and red supergiants located at d$\approx$2.5$\pm$0.4 kpc share a common kinematics and are physically linked.

However, the detailed chemistry of this region is poorly constrained and only a few measurements exist so far. High resolution optical spectra for five  B-type stars members of NGC~457 \citep{NGC457} provided carbon, oxygen, magnesium and silicon abundances, indicating sub-solar values of about $-0.3$~dex, with the only exception of oxygen which resulted enhanced by about 0.2~dex. 

A photometric estimate of about solar metallicity for the h,$\chi$ double cluster has been suggested by \citealt{currie10}.
However, photometric estimates of metallicity in general, and especially in the case of very young stellar populations, should be regarded only as indicative, since isochrone fitting is not sufficiently sensitive to metallicity variations of a few tenths of a dex and it also depends on the adopted age/model. 
For 11 RSGS in the Perseus OB-1 associations, \citet{gazak14}, by using global spectral synthesis of a portion of the J-band spectrum, obtained rather warm temperatures in the 3700-4100 K range and about solar metallicites.

Within TNG Large Program titled SPA - Stellar Population Astrophysics: the detailed, age-resolved chemistry of the Milky Way disk (Program ID A37TAC\_13, PI: L. Origlia), that aims at measuring detailed chemical abundances and radial velocities of the luminous stellar populations of the the Milky Way thin disk and its associated star clusters \citep{Origlia19}, we performed a combined optical and NIR spectroscopic screening at high resolution of the young stellar populations within the Perseus complex.

\begin{figure*}[]
    \label{RSGs_GIANO} 
   \centering
   \includegraphics[scale=1.2]{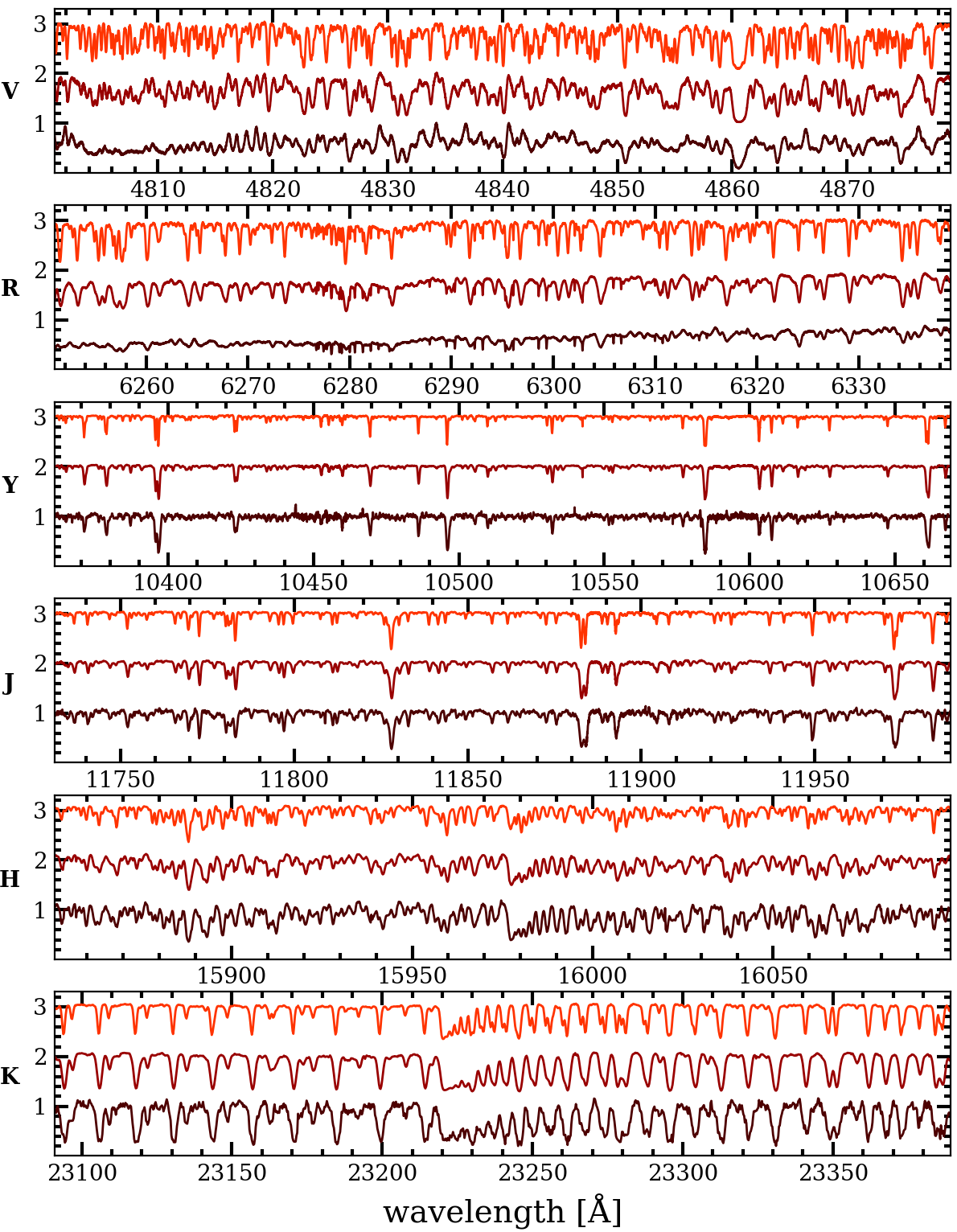} 
   \caption{Portion of the HARPS-N and GIANO-B spectra in the V, R, Y, J, H and K bands for three representative RSGs observed in the Perseus complex. Spectra are corrected for telluric contamination and RV and they are plotted in order of increasing stellar temperature from the bottom to the top. 
   In each panel the normalized spectra of the three stars are shifted by a constant.}
\end{figure*}

\begin{figure*}[h]    
   \centering
   \includegraphics[scale=0.57]{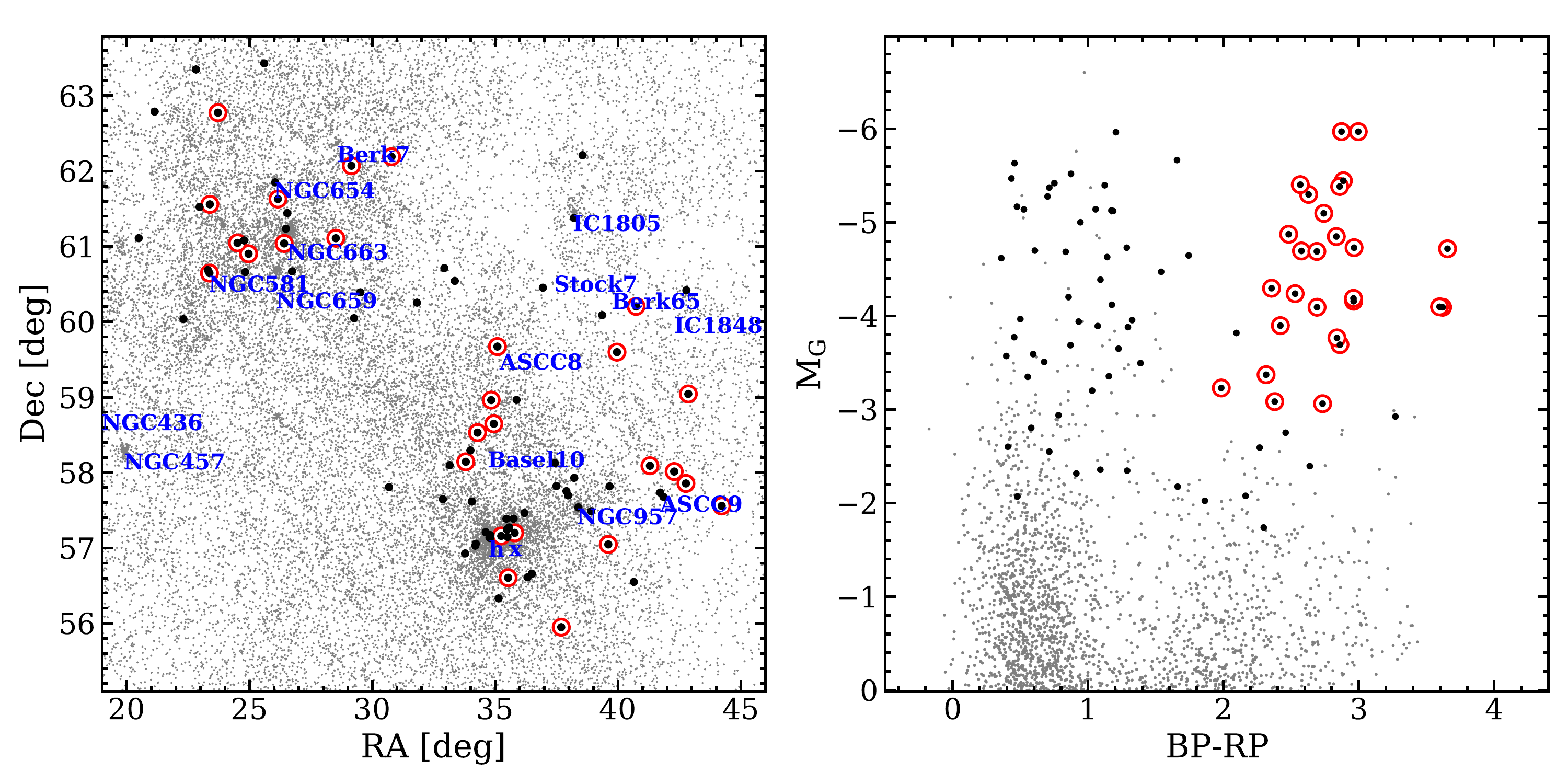}
   \caption{Left panel: RA-Dec map of the stars likely members of the Perseus complex young clusters and associations (grey dots), based on their distances and proper motions, within a projected area of about 10 degrees on sky. Black dots mark the 84 stars spectroscopically observed within the TNG-SPA Large Programme, while red circles mark the 27 RSGs studied in this paper. For reference, we also marked in blue the main clusters in the region.
   Right panel: Gaia eDR3 color magnitude diagram (CMD) for the same stars in the left panel.}
   \label{CMD_diagram}
\end{figure*}

We observed 84 luminous blue and red young stars, candidate members of the Perseus complex, according to their Gaia distances and proper motions. 
A first comprehensive study of the stellar kinematics and spectro-photometric properties from Gaia eDR3 (early data release 3) \citep{gaia_edr3} and SPA observations in the area surrounding $h$ and $\chi$ Per double stellar cluster has been presented in \citet{dalessandro21}.
We found that the region is populated by seven co-moving clusters, that is 
$h$ and $\chi$ Per, NGC~957, Basel 10, N1, N2, N3, the latter three previously unknown, and by an extended and quite massive halo,  defining a complex structure that we named LISCA~I. The kinematic and structural properties of LISCA~I suggest that it might be forming a massive cluster (some $10^5 \textrm M_{\odot}$) through hierarchical assembly. 
Two other kinematics studies of the other young clusters and associations in the Perseus complex are ongoing.

This paper reports the detailed stellar parameters and chemical analysis of a sample of 27 red supergiants (RSGs) distributed over the entire area of the Perseus complex. 
The RSG evolutionary phase for stars with progenitor masses in the 9-25 M${\odot}$ range is quite rapid and mostly characterized by He-core burning at temperatures close to the Hayashi limit. 
Optical and NIR spectra of RSGs are very rich in absorption lines from many   light and heavy metals of interest, enabling  a detailed chemical screening of 
their pristine chemical composition as well as of possible modifications due to mixing processes in their interiors. 
Interesting, 
RSGs hosted in the Perseus complex not only allow to trace the recent star formation history and chemical enrichment of that region, but also to 
investigate their physical, chemical and evolutionary properties in a metallicity regime in between that one of the Solar neighborhood and of the Magellanic Clouds.

Observations and data reduction are described in Sect.~\ref{obs}, spectral analysis and the derived chemical abundances are presented in Sect.~\ref{spec} and Sect.~\ref{chem}, respectively. Discussion of the results and conclusions are reported in Sect.~\ref{disc} and \ref{conclusions}.

\begin{table*}[h]
\centering
\tiny
\addtolength{\tabcolsep}{-1.7pt}    
\caption{Photometric information for the studied RSGs and derived stellar parameters and RVs.}
\label{tab_obs} 
\begin{tabular}{|l|l|cc|rrr|c|c|c||ccc|c|}
\hline
 \# & star & RA(J2000) & DEC(J2000) & \multicolumn{3}{c|}{Gaia EDR3} & 2MASS & E(B-V) & $\rm log\Big(\frac{L_{Bol}}{L_{\odot}}\Big)$  &  $T_{eff}$ & log(g)  & $\xi$ &  RV\\
 & & (h : m : s) &  ($^o$ : ' : '') & G & BP & RP & K & & & [K]  &  [dex]   &  [$\rm km \ s^{-1}$]   & [$\rm km \ s^{-1}$] \\
\hline
\rowcolor{lightgray}
1 & SU Per  & 2:22:06.9 & 56:36:14.9  & 5.93 & 7.70 & 4.70 & 1.46 & 0.36 & 4.95 & 3466 & -0.37 & 3.80 & -49\\
2 & FZ Per & 2:20:59.6 & 57:09:29.9  & 6.86 & 8.30 & 5.67 & 2.48 & 0.51 & 4.64 & 3750 &  0.07 & 3.10 & -44 \\
\rowcolor{lightgray}
3 & V439 Per & 2:23:11.1 & 57:11:57.9  & 7.00 & 8.43 & 5.86 & 2.69 & 0.55 & 4.44 & 3690 &  0.25 & 3.40 & -44  \\
4 & V550 Per & 2:15:13.3 & 58:08:32.4  & 7.59 & 9.10 & 6.41 & 3.19 & 0.74 & 4.62 & 3720 &  0.08 & 3.30 & -47 \\
\rowcolor{lightgray}
5 & PP Per & 2:17:08.2 & 58:31:47.0  & 7.74 & 9.48 & 6.52 & 2.95 & 0.78 & 4.50 & 3670 &  0.18 & 3.00 & -51 \\
6 & BD+57 540 & 2:19:46.9 & 58:38:48.1  & 8.28 & 9.51 & 7.19 & 4.28 & 0.68 & 4.18 & 4125 &  0.70 & 2.10 & -53 \\
\rowcolor{lightgray}
7 & HD 15406 & 2:30:45.7 & 55:56:53.2  & 8.68 & 9.69 & 7.71 & 5.15 & 0.40 & 3.96 & 4190 &  0.95 & 2.00 & -51 \\
8  & YZ Per   & 2:38:25.4 & 57:02:46.2 & 6.45 & 8.17 & 5.28 & 1.91 & 0.62 & 4.85 & 3616 & -0.20 & 4.00 & -43 \\
\rowcolor{lightgray} 
9 & T Per    & 2:19:21.9 & 58:57:40.4 & 7.19 & 8.88 & 5.91 & 1.94 & 0.84 & 4.63 & 3594 &  0.01 & 3.30 & -41 \\
10  & GP Cas & 2:39:50.4 & 59:35:51.3 & 7.45 & 9.76 & 6.10 & 1.95 & 1.07 & 4.74 & 3471 & -0.16 & 3.40 & -41 \\
\rowcolor{lightgray} 
11  & BD+60 299 & 1:39:51.7 & 60:54:08.3 & 6.44 & 8.05 & 5.18 & 2.11 & 0.75 & 5.05 & 3491 & -0.45 & 3.40 & -44 \\
12  & V648 Cas & 2:51:03.9 & 57:51:19.9 & 7.52 & 9.70 & 6.08 & 2.25 & 1.27 & 4.87 & 3572 & -0.24 & 3.50 & -41 \\
\rowcolor{lightgray} 
13 & DO 26429 & 0:56:53.2 & 57:33:24.9 & 7.86 & 10.09 & 6.49 & 2.48 & 1.49 & 4.94 & 3578 & -0.30 & 3.50 & -42 \\
14 & V605 Cas & 2:20:22.5 & 59:40:16.9 & 7.02 & 8.73 & 5.87 & 2.58 & 0.82 & 4.95 & 3736 & -0.24 & 3.10 & -36 \\
\rowcolor{lightgray} 
15  & BD+60 265 & 1:33:33.2 & 61:33:29.7 & 7.23 & 8.79 & 6.05 & 2.63 & 0.88 & 4.61 & 3637 &  0.06 & 3.20 & -41 \\
16  & BD+60 327 & 1:45:38.8 & 61:02:22.8 & 7.08 & 8.49 & 5.93 & 2.93 & 1.26 & 4.58 & 3778 &  0.15 & 3.00 & -46 \\
\rowcolor{lightgray} 
17 & HD 237006 & 2:49:08.8 & 58:00:48.4 & 7.86 & 9.35 & 6.66 & 3.12 & 1.10 & 4.40 & 3863 &  0.37 & 2.80 & -39 \\
18 & BD+59274 & 1:33:29.2 & 60:38:47.7 & 7.44 & 8.79 & 6.31 & 3.12 & 0.51 & 4.62 & 3740 &  0.09 & 2.80 & -45 \\
\rowcolor{lightgray} 
19  & WX Cas   & 1:54:03.7 & 61:06:32.9 & 8.46 & 10.20 & 7.24 & 3.41 & 1.88 & 4.50 & 3787 &  0.23 & 3.00 & -44 \\
20 & IRAS02414+5752 & 2:45:12.2 & 58:05:24.4 & 8.27 & 9.91 & 7.07 & 3.56 & 0.89 & 4.42 & 3850 &  0.34 & 2.60 & -42 \\
\rowcolor{lightgray} 
21  & IRAS01530+6149 & 1:56:35.8 & 62:04:13.2 & 8.19 & 9.81 & 6.97 & 3.74 & 1.04 & 4.49 & 3793 &  0.25 & 2.60 & -45 \\
22  & BD+61 369 & 2:03:08.2 & 62:11:24.0 & 7.85 & 9.25 & 6.72 & 3.78 & 1.06 & 4.24 & 3869 &  0.53 & 2.60 & -44 \\
\rowcolor{lightgray} 
23  & DO 24697 & 1:44:38.3 & 61:37:43.2 & 8.58 & 10.21 & 7.37 & 3.87 & 1.62 & 4.29 & 3896 &  0.50 & 2.40 & -37 \\
24  & BD+60 287 & 1:38:03.6 & 61:02:49.2 & 7.90 & 9.16 & 6.81 & 3.98 & 0.73 & 4.04 & 4009 &  0.79 & 2.50 & -44 \\
\rowcolor{lightgray} 
25  & BD+62 272 & 1:34:52.3 & 62:46:28.8 & 8.42 & 9.73 & 7.31 & 4.20 & 1.18 & 4.34 & 3925 &  0.46 & 2.40 & -40 \\
26  & BD+59 532 & 2:42:56.9 & 60:12:16.1 & 8.80 & 10.35 & 7.62 & 4.20 & 1.08 & 4.14 & 4016 &  0.69 & 2.40 & -44 \\
\rowcolor{lightgray} 
27  & IRAS02476+5850 & 2:51:27.1 & 59:02:34.3 & 9.47 & 10.75 & 8.37 & 5.28 & 0.99 & 3.94 & 4048 &  0.91 & 2.30 & -31 \\
\hline\hline
\end{tabular}
\end{table*}

\section{Observations and data reduction}
\label{obs}

\begin{figure*} 
   \includegraphics[scale=0.74]{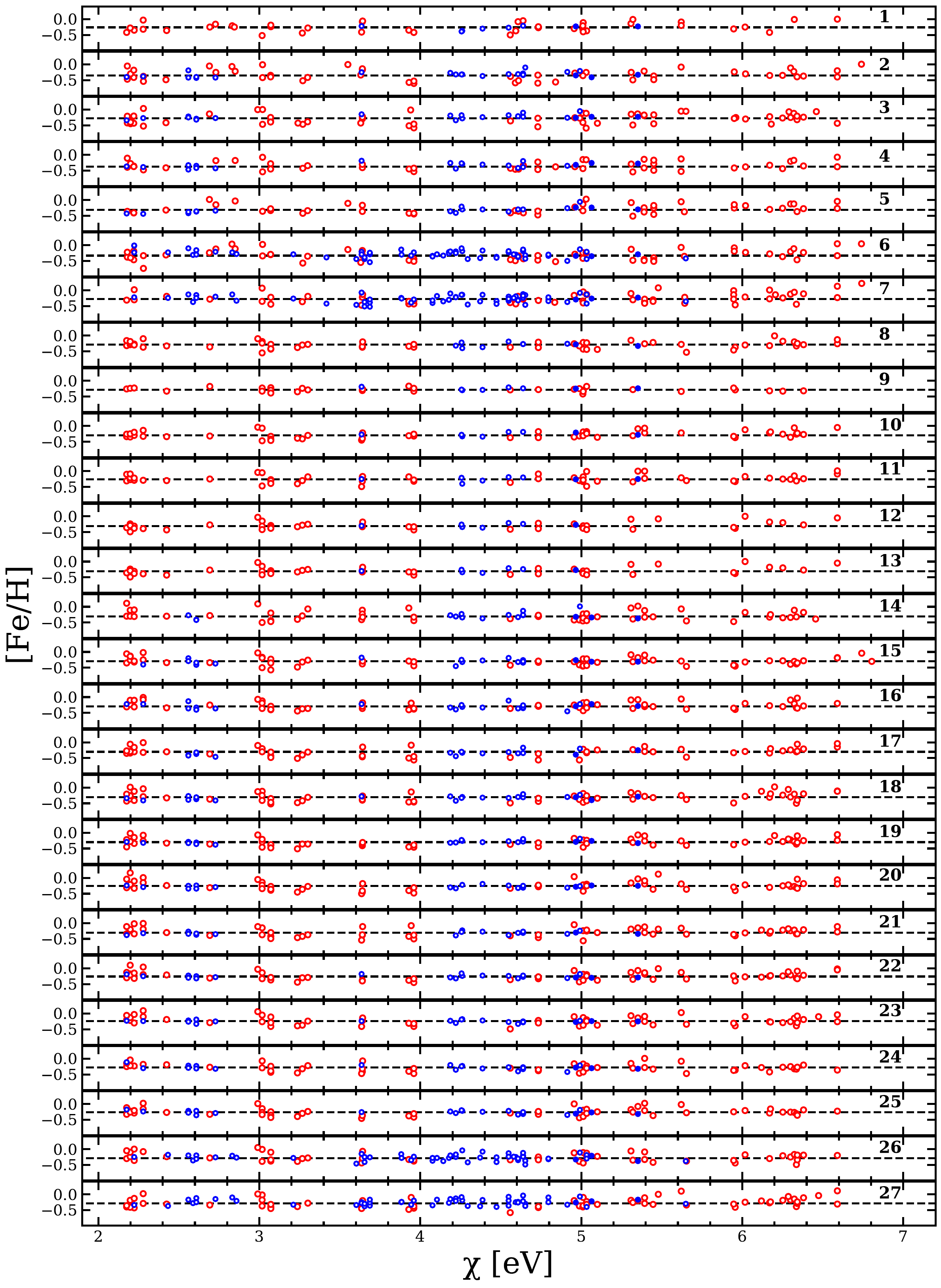}
   \caption{Iron abundances from different optical (blue) and NIR (red) lines as a function of the  excitation potential for the studied RSGs in the Perseus complex. Circles indicates neutral lines, dots single-ionized lines.} 
   \label{Fevschi} 
\end{figure*}

High resolution spectra for 84 young stars in the Persues complex have been simultaneously acquired with the TNG HARPS-N \citep{cosentino14} and GIANO-B \citep{oli12b,oli12a,Origlia14} optical and NIR spectrographs, respectively, on November 3-5, 2018 and on November 6-12, 2019.
HARPS-N covers the optical spectral range ($3780-6910 \ \AA$) at a resolution R$\sim$115,000, while GIANO-B covers the NIR spectral range ($9500-24500 \ \AA$) at a resolution R$\sim$50,000.
The simultaneous observation with HARPS-N and GIANO-B (named the GIARPS configuration) uses a dycroic that splits the optical and NIR light, thus allowing the simultaneous feeding of the two spectrographs \citep{tozzi16,GIARPS17}. On-source integration times between 700s and 3600s, depending on the target brightness, have been set to ensure signal-to-noise ratio $>$50 over the entire spectral range.
GIANO-B spectra are actually the sum of pairs of sub-exposures of 300s obtained by nodding-on-slit, for an optimal subtraction of the background and other instrumental effects. 

HARPS-N spectra were reduced by the instrument Data Reduction Software pipeline and they were normalized using the code {\tt RASSINE}\footnote{\url{https://ascl.net/2102.022}} \citep{rassine20}. The GIANO-B  spectra were reduced using the data reduction pipeline software GOFIO \citep{gofio}, which processes calibration (darks, flats, and U-Ne lamps taken in daytime) and scientific frames. The main feature of GOFIO is the optimal spectral extraction and wavelength calibration based on a physical model of the spectrometer that accurately matches instrumental effects such as variable slit tilt and order curvature over the echellogram \citep{Oli18}. The data reduction package also includes bad pixel and cosmic removal, sky and dark subtraction, flat-field and blaze correction. 
The GIANO-B spectra were also corrected for telluric absorption using the spectra of an O-type standard star taken at different air masses during the same night. The normalised spectra of the telluric standard taken at low and high air mass values were combined with different weights to match the depth of the telluric lines in the target spectra. The spectral regions with telluric transmission lower than 90\% were excluded from the analysis, and the few intrinsic features of the O-stars used as telluric standards are broad lines of H and He that are easily recognizable and corrected using a Voigt fit.
Figure \ref{RSGs_GIANO} shows portions of the observed HARPS-N and the GIANO-B spectra for three representative RSGs here analyzed. 

The RA-Dec map and the Gaia eDR3 colour-magnitude diagram for the stars in the young stellar systems of the Perseus complex  within a projected area of about 10 degrees on sky are shown in Fig.~\ref{CMD_diagram}.  Stars spectroscopically observed within the SPA Large Program are also highlighted. 
Stars have been selected  with distances and proper motions consistent to those of the young clusters and associations in the area. 
Among the 84 observed stars, we classified as "genuine" RSGs the 27 reddest stars in the sample, with M$_G$<-3 or equivalently L$_{bol}>$3.9 L$_{\odot}$.
Three and seven RSGs in our sample are in common with    \citet{gazak14} and \citet{deburgos20} samples, respectively.
The other red stars in their samples were not first priority targets in our list since significantly closer and/or extremely bright in the NIR to be at risk of saturation with GIANO-B (depending on the seeing/cloud conditions) or extremely red.

For the 27 RSGs studied in this paper, Table \ref{tab_obs} lists coordinates, Gaia-eDR3 and 2MASS photometry \citep{2MASS06}, reddening and bolometric luminosity. 
For each star, reddening has been estimated by interpolating the \cite{Schlegel98} extinction maps and applying the corrections by \cite{Schlafly11}. Bolometric luminosities have been estimated by using the de-reddened 2MASS K-band magnitudes, bolometric corrections as prescribed in \citet{Levesque05}.

\section{Spectral Analysis}
\label{spec}

We used spectral synthesis techniques to derive stellar parameters and chemical abundances for the observed 27 RSGs . Synthetic spectra were computed by using the radiative transfer code {\tt TURBOSPECTRUM} \citep{Alvarez&Plez98,Turbospec} with MARCS models atmospheres \citep{MARCS1}. 
For the chemical analysis of the cool RSGs we compiled an optimized list of un-blended and un-saturated atomic and molecular lines, initially selected by using the {\tt TurboSLine} tool described in \citealt{fanelli21} and subsequently confirmed by visual inspection. Optical lines in particular, have been selected in those spectral regions free from significant CN and TiO contamination, by visually inspecting observed and synthetic spectra.
Normally, unsaturated, faint lines do not suffer from major non local thermodynamic equilibrium (NLTE) effects, since they tend to form quite deep in the atmosphere of cool stars.

The importance of using high resolution spectra over a range as wide as possible from the optical to the NIR has been demonstrated in \citealt{fanelli21}, where a combined optical+NIR analysis of Arcturus has been performed. Indeed, a wide spectral coverage has multiple advantages:  {\it i})~it allows to maximize the number of species that can be reliably measured; {\it ii})~it provides multiple diagnostics of stellar parameters; {\it iii})~it enables a number of sanity checks for a self-consistent derivation of both stellar parameters and chemical abundances; {\it iv})~it enables a better physical comprehension of the systematic and degeneracy effects.

For the selected lines the atomic data from VALD3 \citep{Ryabchikova15} and the most updated molecular data from the website of B. Plez \footnote{\url{https://www.lupm.in2p3.fr/users/plez/}} have been used.
Tables A.1-A.3\footnote{Tables A.1-A.3 are only available in electronic form at the CDS via anonymous ftp to ----- or via -----} provide the complete list of optical and NIR atomic and molecular lines used for the chemical abundance analysis. 

The synthetic spectra were convolved with a Gaussian function in order to reproduce the observed line broadening. For the derivation of the chemical abundances we used {\tt SALVADOR}, a tool developed by A. Mucciarelli \citetext{priv.\ comm.} that performs a $\chi ^2$ minimisation between observed and synthetic spectra, and for each line the normalisation of the observed spectra was optimised interactively.

Accurate (at better than 1 km~s$^{-1}$) heliocentric radial velocities for the observed stars have been obtained by means of a standard cross-correlation technique \citep[][]{tonry79} between observed and synthetic spectra with stellar parameters appropriate to fit the studied RSGs.

\begin{figure} 
   \includegraphics[scale=0.49]{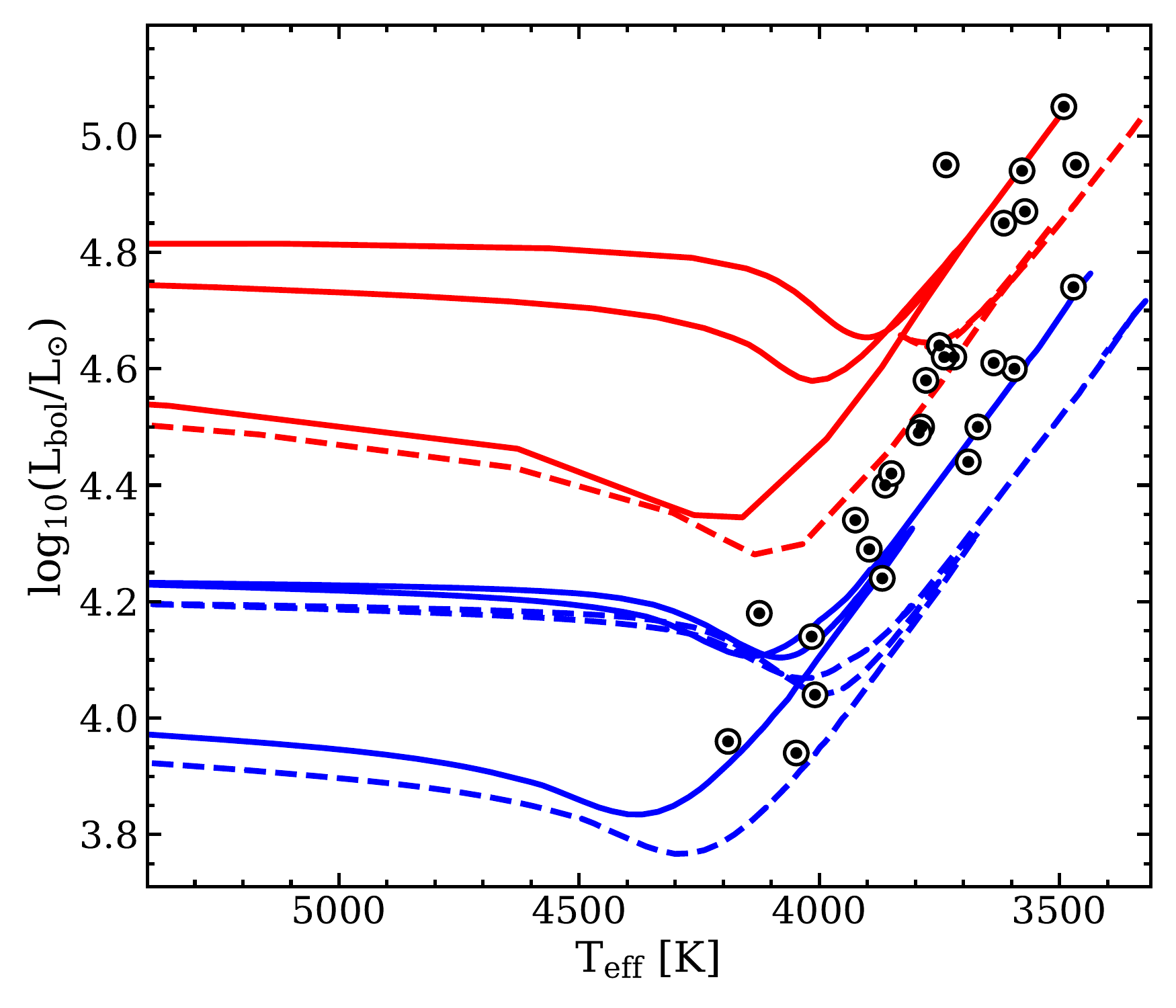}
   \caption{HR diagram for the 27 RSGs studied in this paper and overplotted the evolutionary tracks of 9~M$_{\odot}$ (blue lines) and 14~M$_{\odot}$ (red lines), for half-solar metallicity (solid lines) and solar metallicity (dashed lines) from the PARSEC models \citep[][]{bres12}, for comparison.}
   \label{HR_diagram} 
\end{figure}

\subsection{Stellar parameters}
\label{stel_param}

Stellar parameters for the observed stars are listed in Table~\ref{tab_obs}.
Effective temperatures have been obtained by minimizing the difference between carbon abundances derived from atomic C I and molecular CO lines, the so-called  C-thermometer \citep{fanelli21}, with an average uncertainty of about $80-100$ K and a temperatures range between 3400 and 4200~K.
These spectroscopic temperatures allow also the minimization of any trend between iron abundances and  excitation potentials of both the optical and NIR lines, as shown in Figure~\ref{Fevschi}, 
although this diagnostic is only moderately sensitive at the low RSG temperatures. 

Surface gravities have been obtained by assuming an average stellar mass of 11~M$_{\odot}$, according to their young ages in the $15-30$ Myr range \citep[see][]{currie10,dalessandro21} and using the standard formula 
\begin{equation}\label{logg_formula}
    g = 4 \pi G \sigma_{SF} \frac{M T_{eff}^4}{L_{Bol}}
\end{equation}
where G is the gravitational constant, $\sigma_{SF}$ is the Stefan-Boltzmann constant, 
M is the mass of the star and $L_{Bol}$ is the bolometric luminosity. We found log(g) values ranging from $-0.40$ to $+1.00$ dex, with an overall uncertainty of $\approx \ 0.15$  dex.
A variation of  $\pm$2~M$_{\odot}$ for the adopted mass has little impact (less than 0.1 dex) on the derived gravity.

Microturbulence has been derived by using the standard approach of minimizing the slope between the iron abundance and the reduced equivalent width (EW) of the measured lines, defined as $log_{10}$(EW/$\lambda$). Values in the 2-4~km~s$^{-1}$ range have been obtained, with a typical uncertainty of $\sim$0.2~km~s$^{-1}$.

The line profiles of the observed stars are broader than the instrumental line profile (as determined from telluric lines). This additional broadening is likely due to macroturbulence and it is normally modeled with a Gaussian profile, as for the instrumental broadening. Macroturbulence velocities in the $8-13$ km~s$^{-1}$ range have been inferred. As expected for these evolved stars, we did not find any appreciable line broadening due to stellar rotation. 

The modeling of the post-MS evolution of massive stars is very complex and it depends on the assumed molecular opacities, geometry of the atmospheres and on the used recipes for the treatment of convection, mass loss, rotation etc. as a function of the stellar mass and metallicity \citep[see e.g. discussion in][and references therein]{bertelli09}. 
Although a detailed tuning of model predictions with measured values is beyond the scope of this paper, we take advantage of our homogeneous sample of 27 RSGs in the Perseus complex at half-solar metallicity to attempt a qualitative comparison. 
By using the inferred temperatures and bolometric magnitudes quoted in Table~\ref{tab_obs}, we constructed the Hertzsprung-Russel (HR) diagram of Fig.~\ref{HR_diagram}. The location of the 27 RSGs here analyzed is qualitatively matched by theoretical evolutionary tracks of massive ($\gtrsim$9 $M_{\odot}$) stars, as e.g. those from the PARSEC models \citep[][]{bres12}, 
although it seems that the latter can reach sufficiently high luminosities and low temperatures to match the location of the coolest RSGs only after the He-core burning, that is during the faster phases of shell He-burning  or core C-burning.

\section{Chemical abundances}
\label{chem}
Chemical abundances for 23 species have been derived from more than 200 atomic and molecular lines in the NIR. Their values and corresponding measurement errors are listed in Table~\ref{tab_elem1} and~\ref{tab_elem2}.

Iron abundances were derived from 40-60 NIR lines of Fe~I. The scatter from different lines is typically 0.15~dex. 

\begin{figure*}[!h] 
   \centering
   \includegraphics[scale=0.535]{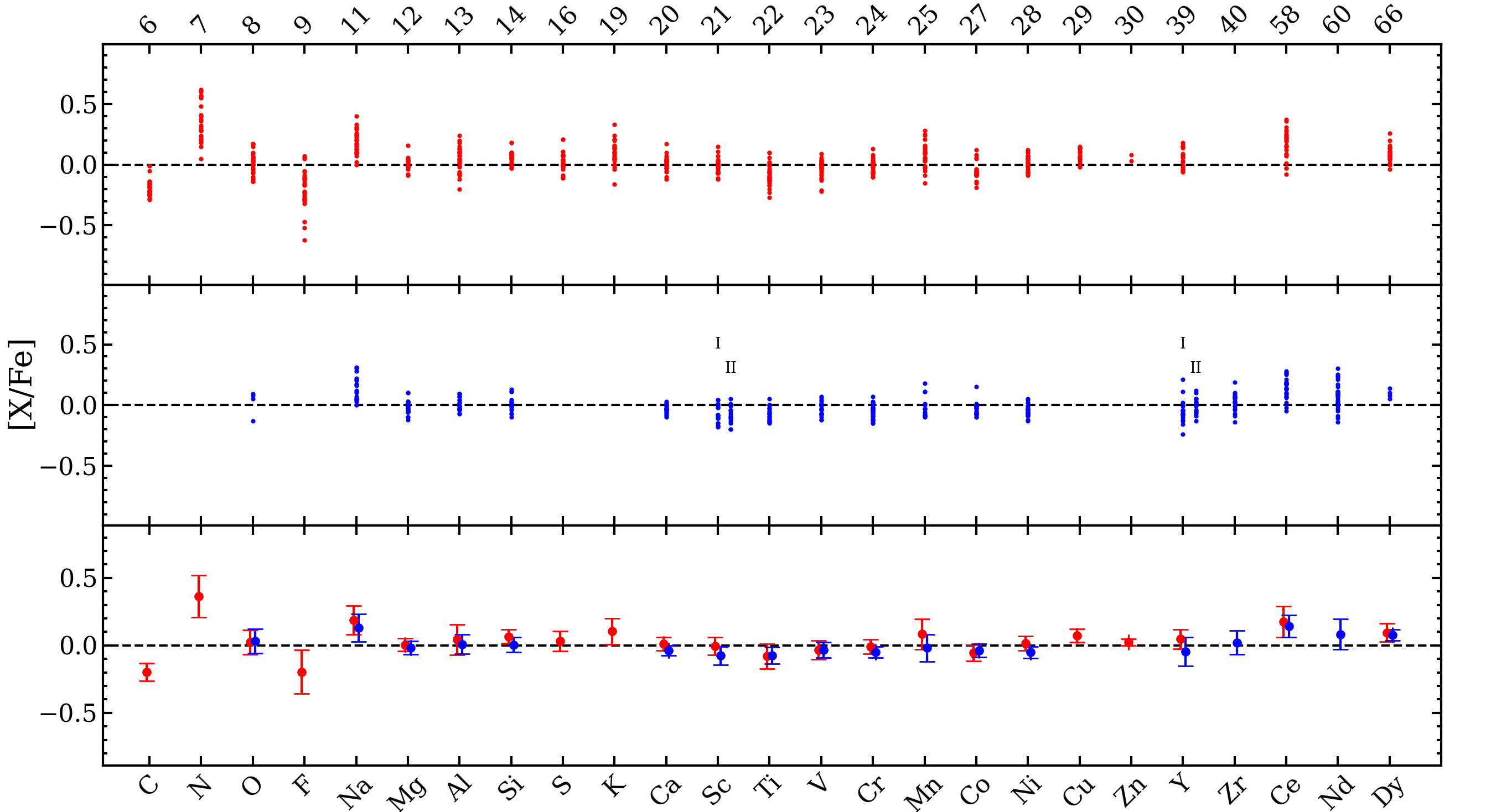}
   \caption{\textit Derived [X/Fe] abundance ratios for the studied RSGs in the Perseus Complex from the NIR (top panel, red symbols) and optical (central panel, blue symbols). Bottom panel shows average [X/Fe] for each element for NIR (red symbols) and optical (blue symbols) with the average errorbars. Values are solar-scaled according to \citet{Grevesse98} solar reference, with the only exception of oxygen for which we used \citet{Asplund09}.}
   \label{abun}
\end{figure*}

From the 27 RSGs here analyzed, an average [Fe/H]=$-0.30\pm0.01$ dex with a dispersion  $\sigma$=0.03 $\pm0.01$~dex has been measured. Concerning the other iron-peak elements,  up to 20 NIR lines of Cr~I could be used to derive  chromium abundance, while a few lines have been used to derived abundances of V~I, Mn~I, Co~I, Ni~I and Cu~I. V~I, Mn~I and Co~I lines have hyperfine structure (HFS).

Zinc abundances could be obtained only for the warmest K-type RSGs from the measurement of two NIR Zn~I lines.
Iron-peak elements homogeneously scale as iron within $\pm 0.1$~dex.
Dozens of unblended lines of Si~I, Ca~I, Ti~I and Mg~I and a few lines of  Na~I, Al~I, S~I, K~I, Sc~I (the latter with HFS) have been measured for each specie, also providing about solar-scaled abundances with the possible exception of Na and K, which turns out to be mildly enhanced with respect to the solar-scaled value. In particular, average [Na/Fe] = +0.18 $\pm$ 0.10 with $\sigma=$ 0.10 $\pm$ 0.01 
and [K/Fe] = +0.10 $\pm$ 0.10 with $\sigma=$ 0.10 $\pm$ 0.01 dex have been obtained.
The measurement of two lines of Y~I and Ce~II and one line of Dy~II (r-process element) provided solar-scaled [Y/Fe], and slightly enhanced [Dy/Fe] and [Ce/Fe] abundance ratios. 
In particular we found average [Dy/Fe] = +0.08 $\pm$ 0.10 with $\sigma=$ 0.09 $\pm$ 0.01 and [Ce/Fe] = +0.17 $\pm$ 0.10 with $\sigma=$ 0.12 $\pm$ 0.01 dex.

CNO abundances have been derived from a few tens of molecular $^{12}$CO, CN and OH lines, respectively.
Following \cite{Ryde09} and \cite{Smith13}, we adopted an iterative method to derive CNO abundances, in order to consider the interplay among these three elements in setting the molecular equilibrium. 
We found depletion of carbon and enhancement of nitrogen with respect to the solar-scaled values, especially in the case of the coolest RSGs, while oxygen abundance turns out to be about solar-scaled. 
In particular,  average [C/Fe]=-0.20$\pm0.01$ dex with a small dispersion of 0.06$\pm$0.01 dex and 
[N/Fe]=+0.36$\pm$0.03 and a dispersion of 0.16$\pm$0.02 dex have been obtained.
It is interesting to notice that the carbon abundance derived from the $\Delta v$=3 $^{12}$CO band-heads in the H-band, is fully consistent with the one derived from individual $^{12}$CO roto-vibrational transitions.
We  used the  $\Delta v$=3 and  $\Delta v$=2 band-heads of $^{13}$CO in the H and K bands, respectively, to also estimate the $^{13}$C abundance and the $^{12}$C/$^{13}$C isotopic abundance ratio, that turns out to be between 20 and 30 in the K-type RSGs and below 12 in the M-type RSGs. 
An average value of $^{12}$C/$^{13}$C = 13.3$\pm$1.2 with a dispersion $\sigma$=6.1$\pm$0.8 has been obtained. \\
Slightly sub-solar values for the fluorine abundance, with an average [F/Fe] = -0.20 $\pm$ 0.10 and $\sigma=$ 0.17 $\pm$ 0.02 dex,  have been obtained by measuring the HF molecular line at $23358.33 \, \AA$ in the K band, and using the transition parameters of \citealt{Jonsson14}. 

Notably, despite the low temperatures and gravities of the analyzed RSGs, chemical abundances for 19 species have been also derived from more than 100 optical atomic lines. \\
Their values and corresponding measurement errors are listed in Table~\ref{tab_elem1} and Table~\ref{tab_elem2}. All the chemical elements measured in the optical are in common with those measured in the NIR with the exception of Zr and Nd, and the derived chemical abundances from the two datasets are fully consistent one to each other within the errors, as also shown in Fig.~\ref{abun}. 

Interestingly, in the optical, iron abundance could be derived from both a few tens neutral and a few single-ionized lines, while oxygen could be measured only in the warmest K giants, where TiO contamination is minimal. Concerning Na~I, we used only the doublet  at $6154.226$ and $6160.747 \ \AA$ since only marginally affected by NLTE effects, with $\le$0.1 dex negative corrections \citep{Mashonkina00,Lind11,Alexeeva14}. 

This doublet provides sodium abundances consistent with those derived from the three selected lines in the NIR at $10746.449 \ \AA$, $16388.850 \ \AA$ (partially blended with a Ni~I line at $16388.729 \ \AA$) and at $21452.373 \ \AA$. 
For the coolest stars, the TiO and CN contamination is so severe that only a few Mg, Fe~I, Fe~II, Cr~I, Ce~II and Nd~II lines could be measured. 

About solar-scaled values for the majority of the measured elements,  including Zr that could not be measured in the NIR, and the slightly enhancement of Na, Y, Dy, Ce and Nd (not measured in the NIR) has been also consistently inferred from the optical lines. 
\begin{landscape}
\begin{table}[t]
\centering
\caption{CNO, F, Na, Al, K and alpha element abundances for the studied RSGs in the Perseus complex. Values are solar-scaled according to \citet{Grevesse98} solar reference, with the only exception of oxygen for which we used \citet{Asplund09}. Numbers in parenthesis represent the number of lines used to determine chemical abundances.}
\scriptsize
\addtolength{\tabcolsep}{-4.5pt}    
\begin{tabular}{|l|c|c|c|cc|c|cc|cc|c||cc|cc|c|cc|cc|c|cc|}
\hline
\# & [C/H] & $^{12}$C/$^{13}$C& [N/H] & \multicolumn{2}{c|}{[O/H]}& [F/H]& \multicolumn{2}{c|}{[Na/H]}& \multicolumn{2}{c|}{[Al/H]} &[K/H] & \multicolumn{2}{c|}{[Mg/H]}& \multicolumn{2}{c|}{[Si/H]}& [S/H] & \multicolumn{2}{c|}{[Ca/H]} & \multicolumn{2}{c|}{[ScI/H]} & [ScII/H] & \multicolumn{2}{c|}{[Ti/H]} \\ 
& NIR & NIR & NIR & NIR & OPT & NIR & NIR & OPT & NIR & OPT & NIR &NIR & OPT& NIR & OPT & NIR & NIR & OPT & NIR & OPT & OPT & NIR & OPT \\
\hline
\rowcolor{lightgray}			
1      & -0.54 (29)  &  7.00 ( 2)  &  0.26 (18)  & -0.33 (17)  &     --      & -0.50 ( 1)  & -0.15 ( 3)  &      --     & -0.33 ( 4)  &      --     & -0.14 ( 1)    & -0.29 ( 4)  & -0.38 ( 2)  & -0.31 (11)  &      --     & -0.14 ( 1)  & -0.32 (10)  &      --     & -0.32 ( 3)   &      --     &      --      & -0.40 (11)  &      --	\\ 
\rowcolor{lightgray} 																 
   &  $\pm$ 0.05 &  $\pm$ 1.50 &  $\pm$ 0.04 &  $\pm$ 0.02 &     --          &  $\pm$ 0.10 &  $\pm$ 0.07 &      --     &  $\pm$ 0.08 &      --     &  $\pm$ 0.10   &  $\pm$ 0.07 &  $\pm$ 0.10 &  $\pm$ 0.04 &      --     &  $\pm$ 0.10 &  $\pm$ 0.06 &      --     &  $\pm$ 0.09  &      --     &      --      &  $\pm$ 0.05 &      --	\\ 
2     & -0.58 (45)  &  8.00 ( 2)  &  0.05 (54)  & -0.35 (34)  &     --       & -0.58 ( 1)  & -0.20 ( 3)  & -0.25 ( 2)  & -0.41 ( 4)  & -0.36 ( 2)  & -0.28 ( 1)    & -0.37 ( 3)  & -0.38 ( 2)  & -0.29 (12)  & -0.39 ( 2)  & -0.36 ( 5)  & -0.31 (14)  & -0.29 ( 3)  & -0.32 ( 5)   & -0.43 ( 4)  &      --      & -0.47 (16)  & -0.35 (17)   \\  
   &  $\pm$ 0.02 &  $\pm$ 1.00 &  $\pm$ 0.02 &  $\pm$ 0.01 &     --          &  $\pm$ 0.10 &  $\pm$ 0.04 &  $\pm$ 0.10 &  $\pm$ 0.07 &  $\pm$ 0.10 &  $\pm$ 0.10   &  $\pm$ 0.08 &  $\pm$ 0.10 &  $\pm$ 0.04 &  $\pm$ 0.10 &  $\pm$ 0.08 &  $\pm$ 0.05 &  $\pm$ 0.07 &  $\pm$ 0.08  &  $\pm$ 0.02 &      --      &  $\pm$ 0.04 &  $\pm$ 0.01	\\ 
\rowcolor{lightgray} 																 
3     & -0.55 (43)  &  8.00 ( 2)  &  0.10 (61)  & -0.37 (31)  &     --       & -0.56 ( 1)  & -0.14 ( 3)  & -0.20 ( 2)  & -0.35 ( 4)  & -0.32 ( 2)  & -0.31 ( 1)    & -0.27 ( 4)  & -0.25 ( 2)  & -0.25 (10)  & -0.25 ( 2)  & -0.37 ( 2)  & -0.28 (13)  & -0.29 ( 5)  & -0.34 ( 4)   & -0.43 ( 3)  &      --      & -0.47 (16)  & -0.35 (17)   \\  
\rowcolor{lightgray} 																 
   &  $\pm$ 0.02 &  $\pm$ 1.00 &  $\pm$ 0.02 &  $\pm$ 0.01 &     --          &  $\pm$ 0.10 &  $\pm$ 0.05 &  $\pm$ 0.10 &  $\pm$ 0.07 &  $\pm$ 0.10 &  $\pm$ 0.10   &  $\pm$ 0.07 &  $\pm$ 0.10 &  $\pm$ 0.04 &  $\pm$ 0.10 &  $\pm$ 0.10 &  $\pm$ 0.04 &  $\pm$ 0.07 &  $\pm$ 0.09  &  $\pm$ 0.02 &      --      &  $\pm$ 0.04 &  $\pm$ 0.01	\\ 
4     & -0.66 (37)  & 11.00 ( 2)  &  0.10 (58)  & -0.34 (16)  &     --       & -0.63 ( 1)  & -0.26 ( 3)  & -0.18 ( 2)  & -0.50 ( 5)  & -0.33 ( 2)  & -0.33 ( 1)    & -0.40 ( 4)  & -0.36 ( 2)  & -0.28 (12)  & -0.33 ( 2)  & -0.31 ( 2)  & -0.28 (12)  & -0.35 ( 5)  & -0.43 ( 6)   &      --     &      --      & -0.44 (15) & -0.42 (25)   \\  
   &  $\pm$ 0.02 &  $\pm$ 1.00 &  $\pm$ 0.01 &  $\pm$ 0.01 &     --          &  $\pm$ 0.10 &  $\pm$ 0.07 &  $\pm$ 0.10 &  $\pm$ 0.05 &  $\pm$ 0.10 &  $\pm$ 0.10   &  $\pm$ 0.04 &  $\pm$ 0.10 &  $\pm$ 0.06 &  $\pm$ 0.10 &  $\pm$ 0.10 &  $\pm$ 0.03 &  $\pm$ 0.04 &  $\pm$ 0.08  &      --     &      --      &  $\pm$ 0.04 &  $\pm$ 0.02	\\ 
\rowcolor{lightgray} 																 
5     & -0.56 (39)  & 10.00 ( 2)  &  0.10 (56)  & -0.34 (13)  &     --       & -0.53 ( 1)  & -0.10 ( 3)  & -0.05 ( 2)  & -0.24 ( 5)  & -0.37 ( 2)  & -0.17 ( 1)    & -0.27 ( 4)  & -0.26 ( 2)  & -0.22 (11)  & -0.25 ( 2)  & -0.33 ( 3)  & -0.25 (12)  & -0.36 ( 5)  & -0.33 ( 3)   & -0.38 ( 4)  &      --      & -0.44 (16)  & -0.36 (22)   \\  
\rowcolor{lightgray} 																 
   &  $\pm$ 0.02 &  $\pm$ 1.00 &  $\pm$ 0.01 &  $\pm$ 0.01 &     --          &  $\pm$ 0.10 &  $\pm$ 0.05 &  $\pm$ 0.10 &  $\pm$ 0.05 &  $\pm$ 0.10 &  $\pm$ 0.10   &  $\pm$ 0.04 &  $\pm$ 0.10 &  $\pm$ 0.03 &  $\pm$ 0.10 &  $\pm$ 0.06 &  $\pm$ 0.03 &  $\pm$ 0.04 &  $\pm$ 0.15  &  $\pm$ 0.04 &      --      &  $\pm$ 0.04 &  $\pm$ 0.02	\\ 
6     & -0.34 (36)  & 29.00 ( 2)  & -0.28 (62)  & -0.23 (13)  & -0.24 ( 1)   & -0.39 ( 1)  & -0.31 ( 3)  & -0.29 ( 2)  & -0.22 ( 5)  & -0.20 ( 2)  & -0.09 ( 1)    & -0.28 ( 7)  & -0.32 ( 2)  & -0.23 (25)  & -0.29 ( 2)  & -0.35 ( 7)  & -0.26 ( 9)  & -0.33 ( 5)  & -0.32 ( 5)   & -0.25 ( 4)  &      --      & -0.32 (28)  & -0.29 (22)   \\  
   &  $\pm$ 0.01 &  $\pm$ 1.50 &  $\pm$ 0.01 &  $\pm$ 0.01 &  $\pm$ 0.10     &  $\pm$ 0.10 &  $\pm$ 0.03 &  $\pm$ 0.10 &  $\pm$ 0.04 &  $\pm$ 0.10 &  $\pm$ 0.10   &  $\pm$ 0.05 &  $\pm$ 0.10 &  $\pm$ 0.03 &  $\pm$ 0.10 &  $\pm$ 0.03 &  $\pm$ 0.04 &  $\pm$ 0.09 &  $\pm$ 0.10  &  $\pm$ 0.07 &      --      &  $\pm$ 0.03 &  $\pm$ 0.03	\\ 
\rowcolor{lightgray} 																 
7     & -0.38 (28)  & 23.00 ( 2)  & -0.18 (56)  & -0.16 (16)  & -0.20 ( 1)   & -0.43 ( 1)  &  0.07 ( 3)  & -0.01 ( 2)  & -0.23 ( 2)  & -0.20 ( 2)  &  0.00 ( 1)    & -0.17 ( 9)  & -0.32 ( 2)  & -0.15 (30)  & -0.29 ( 2)  & -0.34 ( 7)  & -0.16 ( 9)  & -0.33 ( 5)  & -0.26 ( 3)   & -0.25 ( 4)  &      --      & -0.23 (27)  & -0.29 (22)   \\  
\rowcolor{lightgray} 																 
   &  $\pm$ 0.02 &  $\pm$ 1.50 &  $\pm$ 0.01 &  $\pm$ 0.01 &  $\pm$ 0.10     &  $\pm$ 0.10 &  $\pm$ 0.08 &  $\pm$ 0.10 &  $\pm$ 0.10 &  $\pm$ 0.10 &  $\pm$ 0.10   &  $\pm$ 0.02 &  $\pm$ 0.10 &  $\pm$ 0.03 &  $\pm$ 0.10 &  $\pm$ 0.03 &  $\pm$ 0.03 &  $\pm$ 0.11 &  $\pm$ 0.03  &  $\pm$ 0.04 &      --      &  $\pm$ 0.02 &  $\pm$ 0.01	\\ 
8     & -0.55 (31)  &  7.00 ( 2)  & -0.12 (56)  & -0.30 (21)  &     --       & -0.59 ( 1)  & -0.23 ( 3)  & -0.26 ( 3)  & -0.31 ( 3)  & -0.34 ( 3)  & -0.25 ( 1)    & -0.31 ( 3)  & -0.30 ( 2)  & -0.23 (12)  & -0.31 ( 2)  & -0.41 ( 1)  & -0.30 (13)  & -0.40 ( 5)  & -0.41 ( 5)   & -0.47 ( 4)  & -0.37 ( 8)   & -0.43 (28)  & -0.45 (19)   \\  
   &  $\pm$ 0.02 &  $\pm$ 1.00 &  $\pm$ 0.01 &  $\pm$ 0.01 &     --          &  $\pm$ 0.10 &  $\pm$ 0.05 &  $\pm$ 0.08 &  $\pm$ 0.10 &  $\pm$ 0.06 &  $\pm$ 0.10   &  $\pm$ 0.03 &  $\pm$ 0.10 &  $\pm$ 0.04 &  $\pm$ 0.10 &  $\pm$ 0.10 &  $\pm$ 0.05 &  $\pm$ 0.07 &  $\pm$ 0.04  &  $\pm$ 0.03 &  $\pm$ 0.03  &  $\pm$ 0.03 &  $\pm$ 0.02	\\ 
\rowcolor{lightgray} 																 
9    & -0.51 (35)  &  7.00 ( 2)  & -0.08 (58)  & -0.33 (21)  &     --        & -0.59 ( 1)  & -0.29 ( 3)  & -0.33 ( 3)  & -0.28 ( 3)  & -0.34 ( 2)  & -0.23 ( 1)    & -0.28 ( 3)  & -0.30 ( 2)  & -0.21 (11)  & -0.22 ( 1)  & -0.31 ( 2)  & -0.32 (11)  & -0.31 ( 5)  & -0.31 ( 5)   & -0.44 ( 3)  & -0.43 ( 8)   & -0.46 (27)  & -0.45 (17)   \\  
\rowcolor{lightgray} 																 
   &  $\pm$ 0.01 &  $\pm$ 1.50 &  $\pm$ 0.01   &  $\pm$ 0.01 &     --        &  $\pm$ 0.10 &  $\pm$ 0.03 &  $\pm$ 0.08 &  $\pm$ 0.06 &  $\pm$ 0.10 &  $\pm$ 0.10   &  $\pm$ 0.05 &  $\pm$ 0.10 &  $\pm$ 0.03 &  $\pm$ 0.10 &  $\pm$ 0.10 &  $\pm$ 0.03 &  $\pm$ 0.07 &  $\pm$ 0.06  &  $\pm$ 0.03 &  $\pm$ 0.04  &  $\pm$ 0.02 &  $\pm$ 0.05	\\ 
10    & -0.45 (33)  &  9.00 ( 2)  & -0.03 (51) & -0.18 (19)  &     --        & -0.35 ( 1)  & -0.07 ( 3)  & -0.04 ( 2)  & -0.21 ( 1)  & -0.17 ( 3)  & -0.08 ( 1)    & -0.24 ( 3)  & -0.24 ( 2)  & -0.18 (11)  & -0.25 ( 2)  & -0.33 ( 3)  & -0.21 ( 9)  & -0.24 ( 4)  & -0.23 ( 3)   & -0.39 ( 3)  & -0.37 ( 8)   & -0.25 (25)  & -0.31 (19)   \\  
   &  $\pm$ 0.01 &  $\pm$ 1.50 &  $\pm$ 0.01   &  $\pm$ 0.01 &     --        &  $\pm$ 0.10 &  $\pm$ 0.05 &  $\pm$ 0.16 &  $\pm$ 0.10 &  $\pm$ 0.06 &  $\pm$ 0.10   &  $\pm$ 0.05 &  $\pm$ 0.10 &  $\pm$ 0.04 &  $\pm$ 0.10 &  $\pm$ 0.03 &  $\pm$ 0.04 &  $\pm$ 0.09 &  $\pm$ 0.03  &  $\pm$ 0.04 &  $\pm$ 0.03  &  $\pm$ 0.02 &  $\pm$ 0.02	\\ 
\rowcolor{lightgray} 																 
11    & -0.49 (31)  &  6.00 ( 2)  & -0.06 (57) & -0.13 (21)  &     --        & -0.25 ( 1)  & -0.10 ( 3)  & -0.20 ( 1)  & -0.22 ( 2)  & -0.22 ( 3)  & -0.09 ( 1)    & -0.27 ( 3)  & -0.30 ( 2)  & -0.31 (12)  & -0.31 ( 1)  & -0.30 ( 3)  & -0.30 (11)  & -0.39 ( 4)  & -0.15 ( 3)   & -0.32 ( 2)  & -0.46 ( 6)   & -0.20 (25)  & -0.26 (16)   \\  
\rowcolor{lightgray} 																 
   &  $\pm$ 0.01 &  $\pm$ 1.50 &  $\pm$ 0.01   &  $\pm$ 0.01 &     --        &  $\pm$ 0.10 &  $\pm$ 0.03 &  $\pm$ 0.10 &  $\pm$ 0.10 &  $\pm$ 0.06 &  $\pm$ 0.10   &  $\pm$ 0.08 &  $\pm$ 0.10 &  $\pm$ 0.03 &  $\pm$ 0.10 &  $\pm$ 0.03 &  $\pm$ 0.03 &  $\pm$ 0.09 &  $\pm$ 0.05  &  $\pm$ 0.10 &  $\pm$ 0.08  &  $\pm$ 0.03 &  $\pm$ 0.03	\\ 
12   & -0.48 (28)  &  9.00 ( 2)  &  0.02 (54)  & -0.20 (21)  &     --        & -0.38 ( 1)  & -0.04 ( 3)  & -0.13 ( 2)  & -0.21 ( 3)  & -0.23 ( 3)  & -0.12 ( 1)    & -0.28 ( 3)  & -0.30 ( 2)  & -0.29 (12)  & -0.35 ( 2)  & -0.18 ( 3)  & -0.27 (11)  & -0.34 ( 5)  & -0.27 ( 3)   & -0.35 ( 4)  & -0.36 ( 8)   & -0.33 (25)  & -0.31 (20)   \\  
   &  $\pm$ 0.01 &  $\pm$ 1.00 &  $\pm$ 0.02   &  $\pm$ 0.01 &     --        &  $\pm$ 0.10 &  $\pm$ 0.03 &  $\pm$ 0.08 &  $\pm$ 0.06 &  $\pm$ 0.06 &  $\pm$ 0.10    &  $\pm$ 0.05 &  $\pm$ 0.10 &  $\pm$ 0.03 &  $\pm$ 0.10 &  $\pm$ 0.02 &  $\pm$ 0.02 &  $\pm$ 0.09 &  $\pm$ 0.04  &  $\pm$ 0.04 &  $\pm$ 0.03  &  $\pm$ 0.02 &  $\pm$ 0.02	\\ 
\rowcolor{lightgray} 																 
13   & -0.50 (28)  &  8.00 ( 2)  &  0.02 (55)  & -0.25 (20)  &     --        & -0.31 ( 1)  & -0.02 ( 3)  & -0.04 ( 1)  & -0.33 ( 1)  & -0.27 ( 2)  & -0.15 ( 1)    & -0.28 ( 3)  & -0.30 ( 2)  & -0.25 (10)  & -0.27 ( 2)  & -0.25 ( 3)  & -0.24 (11)  & -0.28 ( 5)  & -0.27 ( 4)   & -0.42 ( 4)  & -0.45 ( 8)   & -0.36 (23)  & -0.38 (22)   \\  
\rowcolor{lightgray} 																 
   &  $\pm$ 0.01 &  $\pm$ 1.00 &  $\pm$ 0.02   &  $\pm$ 0.01 &     --        &  $\pm$ 0.10 &  $\pm$ 0.01 &  $\pm$ 0.10 &  $\pm$ 0.10 &  $\pm$ 0.10 &  $\pm$ 0.10   &  $\pm$ 0.08 &  $\pm$ 0.10 &  $\pm$ 0.03 &  $\pm$ 0.10 &  $\pm$ 0.03 &  $\pm$ 0.03 &  $\pm$ 0.05 &  $\pm$ 0.03  &  $\pm$ 0.03 &  $\pm$ 0.03  &  $\pm$ 0.02 &  $\pm$ 0.02	\\ 
14   & -0.49 (38)  &  9.00 ( 2)  &  0.01 (55)  & -0.24 (21)  &     --        & -0.42 ( 1)  & -0.22 ( 3)  & -0.29 ( 2)  & -0.33 ( 3)  & -0.36 ( 2)  & -0.17 ( 1)    & -0.32 ( 3)  & -0.30 ( 2)  & -0.25 (14)  & -0.28 ( 1)  & -0.24 ( 5)  & -0.27 (12)  & -0.34 ( 2)  & -0.32 ( 3)   & -0.41 ( 4)  & -0.52 ( 5)   & -0.35 (32)  & -0.39 (17)   \\  
   &  $\pm$ 0.01 &  $\pm$ 1.00 &  $\pm$ 0.02   &  $\pm$ 0.01 &     --        &  $\pm$ 0.10 &  $\pm$ 0.03 &  $\pm$ 0.10 &  $\pm$ 0.08 &  $\pm$ 0.10 &  $\pm$ 0.10   &  $\pm$ 0.04 &  $\pm$ 0.10 &  $\pm$ 0.05 &  $\pm$ 0.10 &  $\pm$ 0.08 &  $\pm$ 0.02 &  $\pm$ 0.10 &  $\pm$ 0.05  &  $\pm$ 0.01 &  $\pm$ 0.08  &  $\pm$ 0.02 &  $\pm$ 0.05	\\ 
\rowcolor{lightgray} 																 
15    & -0.42 (27)  & 16.00 ( 2)  & -0.09 (57)  & -0.13 (17)  &     --       & -0.21 ( 1)  &  0.01 ( 3)  & -0.05 ( 2)  & -0.15 ( 1)  & -0.23 ( 3)  & -0.08 ( 1)    & -0.29 ( 3)  & -0.28 ( 2)  & -0.28 (12)  & -0.27 ( 2)  & -0.21 ( 3)  & -0.25 (10)  & -0.30 ( 5)  & -0.17 ( 3)   & -0.29 ( 3)  & -0.32 ( 8)   & -0.22 (23)  & -0.32 (15)   \\  
\rowcolor{lightgray} 																 
   &  $\pm$ 0.01 &  $\pm$ 1.00 &  $\pm$ 0.02    &  $\pm$ 0.01 &     --       &  $\pm$ 0.10 &  $\pm$ 0.03 &  $\pm$ 0.10 &  $\pm$ 0.10 &  $\pm$ 0.06 &  $\pm$ 0.10   &  $\pm$ 0.05 &  $\pm$ 0.10 &  $\pm$ 0.05 &  $\pm$ 0.10 &  $\pm$ 0.06 &  $\pm$ 0.03 &  $\pm$ 0.08 &  $\pm$ 0.03  &  $\pm$ 0.08 &  $\pm$ 0.04  &  $\pm$ 0.02 &  $\pm$ 0.03	\\ 
16    & -0.55 (27)  & 22.00 ( 2)  &  0.30 (26)  & -0.38 (21)  &     --       & -0.88 ( 1)  & -0.02 ( 3)  &     --      & -0.35 ( 3)  &     --      & -0.21 ( 1)    & -0.23 ( 3)  & -0.29 ( 2)  & -0.08 (13)  &     --      & -0.19 ( 2)  & -0.26 (12)  &     --      & -0.38 ( 4)   &     --      &      --      & -0.53 (29)  &     --      \\  
   &  $\pm$ 0.02 &  $\pm$ 1.00 &  $\pm$ 0.03    &  $\pm$ 0.01 &     --       &  $\pm$ 0.10 &  $\pm$ 0.09 &     --      &  $\pm$ 0.08 &     --      &  $\pm$ 0.10   &  $\pm$ 0.07 &  $\pm$ 0.10 &  $\pm$ 0.06 &     --      &  $\pm$ 0.10 &  $\pm$ 0.04 &     --      &  $\pm$ 0.09  &     --      &      --      &  $\pm$ 0.02 &     --	\\ 
\rowcolor{lightgray} 																 
17   & -0.52 (34)  & 17.00 ( 2)  &  0.10 (38)   & -0.36 (21)  &     --       & -0.77 ( 1)  & -0.14 ( 3)  & -0.15 ( 1)  & -0.38 ( 3)  & -0.39 ( 2)  & -0.32 ( 1)    & -0.32 ( 3)  & -0.29 ( 2)  & -0.21 (12)  & -0.29 ( 2)  & -0.09 ( 3)  & -0.36 (13)  & -0.35 ( 5)  & -0.37 ( 5)   & -0.47 ( 2)  & -0.41 ( 5)   & -0.45 (28)  & -0.46 (18)   \\  
\rowcolor{lightgray} 																 
   &  $\pm$ 0.02 &  $\pm$ 1.00 &  $\pm$ 0.02 &  $\pm$ 0.01    &     --       &  $\pm$ 0.10 &  $\pm$ 0.02 &  $\pm$ 0.10 &  $\pm$ 0.06 &  $\pm$ 0.10 &  $\pm$ 0.10   &  $\pm$ 0.02 &  $\pm$ 0.10 &  $\pm$ 0.06 &  $\pm$ 0.10 &  $\pm$ 0.10 &  $\pm$ 0.04 &  $\pm$ 0.10 &  $\pm$ 0.06  &  $\pm$ 0.10 &  $\pm$ 0.04  &  $\pm$ 0.01 &  $\pm$ 0.01	\\ 
18    & -0.45 (33)  & 10.00 ( 2)  &  0.02 (58)  & -0.23 (19)  &     --       & -0.53 ( 1)  & -0.18 ( 2)  & -0.21 ( 3)  & -0.16 ( 3)  & -0.19 ( 3)  & -0.16 ( 1)    & -0.26 ( 3)  & -0.16 ( 2)  & -0.19 (13)  & -0.13 ( 2)  & -0.26 ( 5)  & -0.27 (12)  & -0.30 ( 4)  & -0.35 ( 5)   & -0.34 ( 4)  & -0.21 ( 8)   & -0.38 (29)  & -0.38 (21)   \\  
   &  $\pm$ 0.01 &  $\pm$ 1.00 &  $\pm$ 0.01 &  $\pm$ 0.01    &     --       &  $\pm$ 0.10 &  $\pm$ 0.10 &  $\pm$ 0.08 &  $\pm$ 0.04 &  $\pm$ 0.05 &  $\pm$ 0.10   &  $\pm$ 0.07 &  $\pm$ 0.10 &  $\pm$ 0.08 &  $\pm$ 0.10 &  $\pm$ 0.06 &  $\pm$ 0.04 &  $\pm$ 0.05 &  $\pm$ 0.03  &  $\pm$ 0.03 &  $\pm$ 0.10  &  $\pm$ 0.02 &  $\pm$ 0.01	\\ 
\rowcolor{lightgray} 																 
19    & -0.53 (30)  & 12.00 ( 2)  &  0.28 (43)  & -0.43 (20)  &     --       & -0.81 ( 1)  &  0.02 ( 1)  &     --      & -0.49 ( 3)  &     --      & -0.45 ( 1)    & -0.38 ( 2)  & -0.31 ( 1)  & -0.24 ( 9)  &     --      & -0.25 ( 3)  & -0.41 ( 7)  &     --      & -0.29 ( 4)   &     --      &      --      & -0.52 (19)  &     --      \\  
\rowcolor{lightgray} 																 
   &  $\pm$ 0.01 &  $\pm$ 1.00 &  $\pm$ 0.03    &  $\pm$ 0.01 &     --       &  $\pm$ 0.10 &  $\pm$ 0.10 &     --      &  $\pm$ 0.08 &     --      &  $\pm$ 0.10   &  $\pm$ 0.10 &  $\pm$ 0.10 &  $\pm$ 0.04 &     --      &  $\pm$ 0.04 &  $\pm$ 0.03 &     --      &  $\pm$ 0.05  &     --      &      --      &  $\pm$ 0.02 &     --	\\ 
20   & -0.47 (26)  &  9.00 ( 2)  &  0.03 (51)   & -0.22 (21)  & -0.18 ( 1)   & -0.42 ( 1)  &  0.08 ( 1)  &  0.04 ( 2)  & -0.15 ( 2)  & -0.22 ( 3)  & -0.09 ( 1)    & -0.29 ( 3)  & -0.31 ( 2)  & -0.28 (10)  & -0.33 ( 2)  & -0.24 ( 3)  & -0.28 ( 9)  & -0.32 ( 4)  & -0.22 ( 3)   & -0.26 ( 4)  & -0.30 ( 7)   & -0.25 (26)  & -0.28 (14)   \\  
   &  $\pm$ 0.01 &  $\pm$ 1.00 &  $\pm$ 0.01    & $\pm$ 0.01  &  $\pm$ 0.10  &  $\pm$ 0.10 &  $\pm$ 0.10 &  $\pm$ 0.10 &  $\pm$ 0.10 &  $\pm$ 0.08 &  $\pm$ 0.10   &  $\pm$ 0.06 &  $\pm$ 0.10 &  $\pm$ 0.05 &  $\pm$ 0.10 &  $\pm$ 0.02 &  $\pm$ 0.02 &  $\pm$ 0.08 &  $\pm$ 0.04  &  $\pm$ 0.04 &  $\pm$ 0.10  &  $\pm$ 0.02 &  $\pm$ 0.04	\\ 
\rowcolor{lightgray} 																
21    & -0.43 (27)  & 13.00 ( 2)  &  0.28 (28)  & -0.26 (21)  &     --       & -0.56 ( 1)  & -0.04 ( 1)  &     --      & -0.09 ( 2)  &     --      & -0.24 ( 1)    & -0.28 ( 3)  & -0.36 ( 2)  & -0.21 (13)  &     --      & -0.27 ( 2)  & -0.32 (10)  &     --      & -0.35 ( 2)   &     --      &      --      & -0.46 (30)  &     --      \\  
\rowcolor{lightgray} 																 
   &  $\pm$ 0.02 &  $\pm$ 1.00 &  $\pm$ 0.03    &  $\pm$ 0.01 &     --       &  $\pm$ 0.10 &  $\pm$ 0.10 &     --      &  $\pm$ 0.10 &     --      &  $\pm$ 0.10   &  $\pm$ 0.08 &  $\pm$ 0.10 &  $\pm$ 0.08 &     --      &  $\pm$ 0.10 &  $\pm$ 0.04 &     --      &  $\pm$ 0.10  &     --      &      --      &  $\pm$ 0.02 &     --	\\ 
22    & -0.46 (28)  & 17.00 ( 2)  & -0.01 (56)  & -0.13 (16)  &     --       & -0.24 ( 1)  &  0.00 ( 1)  & -0.18 ( 1)  & -0.25 ( 2)  & -0.37 ( 2)  & -0.20 ( 1)    & -0.31 ( 3)  & -0.31 ( 2)  & -0.31 (10)  & -0.35 ( 2)  & -0.34 ( 3)  & -0.28 (12)  & -0.35 ( 3)  & -0.26 ( 1)   & -0.33 ( 3)  & -0.33 ( 4)   & -0.28 (26)  & -0.43 (10)   \\  
   &  $\pm$ 0.02 &  $\pm$ 1.00 &  $\pm$ 0.01    &  $\pm$ 0.01 &     --       &  $\pm$ 0.10 &  $\pm$ 0.10 &  $\pm$ 0.10 &  $\pm$ 0.10 &  $\pm$ 0.10 &  $\pm$ 0.10   &  $\pm$ 0.06 &  $\pm$ 0.10 &  $\pm$ 0.03 &  $\pm$ 0.10 &  $\pm$ 0.03 &  $\pm$ 0.04 &  $\pm$ 0.10 &  $\pm$ 0.10  &  $\pm$ 0.02 &  $\pm$ 0.08  &  $\pm$ 0.02 &  $\pm$ 0.02	\\ 
\rowcolor{lightgray} 																 
23    & -0.47 (37)  & 19.00 ( 2)  &  0.25 (26)  & -0.37 (21)  &     --       & -0.46 ( 1)  & -0.04 ( 1)  &     --      & -0.15 ( 3)  &     --      & -0.27 ( 1)    & -0.27 ( 3)  & -0.30 ( 1)  & -0.22 (12)  &     --      & -0.19 ( 2)  & -0.26 (11)  &     --      & -0.34 ( 2)   &     --      &      --      & -0.44 (28)  &     --      \\  
\rowcolor{lightgray} 																 
   &  $\pm$ 0.01 &  $\pm$ 1.00 &  $\pm$ 0.02    &  $\pm$ 0.01 &     --       &  $\pm$ 0.10 &  $\pm$ 0.10 &     --      &  $\pm$ 0.01 &     --      &  $\pm$ 0.10   &  $\pm$ 0.07 &  $\pm$ 0.10 &  $\pm$ 0.04 &     --      &  $\pm$ 0.10 &  $\pm$ 0.04 &     --      &  $\pm$ 0.10  &     --      &      --      &  $\pm$ 0.02 &     --	\\ 
24    & -0.43 (23)  & 21.00 ( 2)  & -0.05 (57)  & -0.19 (19)  &     --       & -0.41 ( 1)  & -0.18 ( 1)  & -0.16 ( 3)  & -0.10 ( 4)  & -0.19 ( 3)  & -0.14 ( 1)    & -0.30 ( 2)  & -0.27 ( 2)  & -0.19 (12)  & -0.23 ( 2)  & -0.24 ( 2)  & -0.28 (11)  & -0.32 ( 5)  & -0.26 ( 5)   & -0.34 ( 4)  & -0.27 ( 7)   & -0.34 (27)  & -0.37 (22)   \\  
   &  $\pm$ 0.02 &  $\pm$ 1.00 &  $\pm$ 0.01    &  $\pm$ 0.01 &     --       &  $\pm$ 0.10 &  $\pm$ 0.10 &  $\pm$ 0.06 &  $\pm$ 0.01 &  $\pm$ 0.06 &  $\pm$ 0.10   &  $\pm$ 0.10 &  $\pm$ 0.10 &  $\pm$ 0.04 &  $\pm$ 0.10 &  $\pm$ 0.10 &  $\pm$ 0.02 &  $\pm$ 0.08 &  $\pm$ 0.07  &  $\pm$ 0.03 &  $\pm$ 0.04  &  $\pm$ 0.02 &  $\pm$ 0.02	\\ 
\rowcolor{lightgray} 																 
25    & -0.50 (20)  & 14.00 ( 2)  &  0.29 (15)  & -0.27 (18)  &     --       & -0.40 ( 1)  & -0.14 ( 1)  &     --      & -0.07 ( 1)  &     --      & -0.21 ( 1)    & -0.33 ( 1)  & -0.34 ( 1)  & -0.23 ( 7)  &     --      &	   --     & -0.30 ( 7)  &     --      & -0.34 ( 1)   &     --     &      --      & -0.41 (22)  &     --      \\  
\rowcolor{lightgray} 																 
   &  $\pm$ 0.02 &  $\pm$ 1.00 &  $\pm$ 0.04    &  $\pm$ 0.02 &     --       &  $\pm$ 0.10 &  $\pm$ 0.10 &     --      &  $\pm$ 0.10 &     --      &  $\pm$ 0.10   &  $\pm$ 0.10 &  $\pm$ 0.10 &  $\pm$ 0.04 &     --      &      --     &  $\pm$ 0.06 &     --      &  $\pm$ 0.10  &     --      &      --      &  $\pm$ 0.03 &     --	\\ 
26    & -0.56 (24)  & 19.00 ( 2)  &  0.31 (18)  & -0.45 (17)  & -0.41 ( 1)   & -0.59 ( 1)  & -0.21 ( 3)  & -0.24 ( 2)  & -0.27 ( 3)  & -0.31 ( 2)  & -0.31 ( 1)    & -0.33 ( 3)  & -0.40 ( 2)  & -0.21 ( 8)  & -0.32 ( 1)  & -0.27 ( 2)  & -0.36 (10)  & -0.35 ( 2)  & -0.38 ( 4)   & -0.39 ( 2)  & -0.39 ( 5)   & -0.42 (25)  & -0.42 (12)   \\  
   &  $\pm$ 0.02 &  $\pm$ 1.00 &  $\pm$ 0.04    &  $\pm$ 0.01 &  $\pm$ 0.10  &  $\pm$ 0.10 &  $\pm$ 0.01 &  $\pm$ 0.10 &  $\pm$ 0.05 &  $\pm$ 0.10 &  $\pm$ 0.10   &  $\pm$ 0.04 &  $\pm$ 0.10 &  $\pm$ 0.02 &  $\pm$ 0.10 &  $\pm$ 0.10 &  $\pm$ 0.02 &  $\pm$ 0.10 &  $\pm$ 0.05  &  $\pm$ 0.10 &  $\pm$ 0.03  &  $\pm$ 0.03 &  $\pm$ 0.04	\\ 
\rowcolor{lightgray} 																 
27    & -0.53 (23)  & 19.00 ( 2)  &  0.08 (27)  & -0.28 (19)  &     --       & -0.60 ( 1)  & -0.15 ( 1)  &     --      & -0.17 ( 1)  &     --      & -0.20 ( 1)    & -0.36 ( 2)  & -0.36 ( 2)  & -0.23 ( 9)  &     --      & -0.20 ( 1)  & -0.38 ( 9)  &     --      & -0.28 ( 2)   &     --      &      --      & -0.40 (22)  &     --      \\  
\rowcolor{lightgray} 																 
   &  $\pm$ 0.02 &  $\pm$ 1.50   &  $\pm$ 0.03  &  $\pm$ 0.01 &     --       &  $\pm$ 0.10 &  $\pm$ 0.10 &     --      &  $\pm$ 0.10 &     --      &  $\pm$ 0.10   &  $\pm$ 0.10 &  $\pm$ 0.10 &  $\pm$ 0.03 &     --      &  $\pm$ 0.10 &  $\pm$ 0.02 &     --      &  $\pm$ 0.10  &     --      &      --      &  $\pm$ 0.03 &     --	\\ 
\hline\hline
\end{tabular}
\label{tab_elem1}
\end{table}
\end{landscape}

\begin{landscape}
\begin{table}[ht]
\centering
\caption{Iron, iron-peak and neutron-capture element abundances for the studied RSGs in the Perseus complex. Values are solar-scaled according to \citet{Grevesse98} solar reference. Numbers in parenthesis represent the number of lines used to determine chemical abundances}
\scriptsize
\addtolength{\tabcolsep}{-5.2pt}
\begin{tabular} {|c|cc|c|cc|cc|cc|cc|cc|c|c|cc|c|c|cc|c|cc|}
\hline 
\# & \multicolumn{2}{c|}{[FeI/H]} & [FeII/H] & \multicolumn{2}{c|}{[V/H]}& \multicolumn{2}{c|}{[Cr/H]} & \multicolumn{2}{c|}{[Mn/H]}&\multicolumn{2}{c|}{[Co/H]}& \multicolumn{2}{c|}{[Ni/H]}&[Cu/H] & [Zn/H]& \multicolumn{2}{c|}{[YI/H]} & [YII/H] & [Zr/H]&\multicolumn{2}{c|}{[Ce/H]}& [Nd/Fe] & \multicolumn{2}{c|}{[Dy/H]}\\
& NIR & OPT & OPT & NIR & OPT & NIR & OPT & NIR & OPT & NIR & OPT& NIR & OPT & NIR &NIR&NIR & OPT & OPT & OPT & NIR & OPT & OPT &  NIR & OPT \\
\hline
\rowcolor{lightgray} 							
1     & -0.35 (42)   & -0.32 ( 6)   & -0.35 ( 2)  & -0.26 ( 3)  &      --      & -0.27 (19)   & -0.30 ( 1)  & -0.25 ( 2)  &      --     & -0.23 ( 4)  &      --      & -0.28 ( 4)  &	   --     & -0.24 ( 1)   &      --     & -0.21 ( 1)  &      --      &      --      &	  --	  &	 --	         & -0.15 ( 1)  & -0.23 ( 2)  &      --      &	  --	   \\ 
\rowcolor{lightgray} 																
      &  $\pm$ 0.03  &  $\pm$ 0.04  &  $\pm$ 0.10 &  $\pm$ 0.04 &      --      &  $\pm$ 0.04  &  $\pm$ 0.10 &  $\pm$ 0.10 &      --     &  $\pm$ 0.06 &      --      &  $\pm$ 0.09 &      --      &  $\pm$ 0.10  &      --     &  $\pm$ 0.10 &      --      &      --      &	  --	  &	 --	         &  $\pm$ 0.10 &  $\pm$ 0.10 &      --      &	  --	   \\ 
2     & -0.35 (52)   & -0.32 (23)   & -0.35 ( 3)  & -0.36 ( 3)  & -0.26 (14)   & -0.36 (18)   & -0.30 ( 9)  & -0.37 ( 2)  & -0.38 ( 1)  & -0.44 ( 3)  & -0.33 ( 7)   & -0.39 ( 6)  & -0.36 ( 7)   & -0.21 ( 1)   &      --     & -0.21 ( 2)  &      --      &      --      & -0.30 ( 7)   & -0.11 ( 2)   & -0.15 ( 2)  & -0.23 ( 8)  & -0.26 ( 1)   & -0.22 ( 1)   \\ 
      &  $\pm$ 0.02  &  $\pm$ 0.02  &  $\pm$ 0.10 &  $\pm$ 0.03 &  $\pm$ 0.01  &  $\pm$ 0.04  &  $\pm$ 0.03 &  $\pm$ 0.10 &  $\pm$ 0.10 &  $\pm$ 0.08 &  $\pm$ 0.02  &  $\pm$ 0.12 &  $\pm$ 0.02  &  $\pm$ 0.10  &      --     &  $\pm$ 0.10 &      --      &      --      &  $\pm$ 0.03  &  $\pm$ 0.10  &  $\pm$ 0.10 &  $\pm$ 0.04 &  $\pm$ 0.10  &  $\pm$ 0.10  \\ 
\rowcolor{lightgray} 																
3     & -0.27 (61)   & -0.25 (23)   & -0.23 ( 3)  & -0.39 ( 3)  & -0.26 (14)   & -0.37 (18)   & -0.30 ( 9)  & -0.42 ( 2)  & -0.32 ( 1)  & -0.36 ( 3)  & -0.25 ( 8)   & -0.25 ( 7)  & -0.23 ( 7)   & -0.25 ( 1)   &      --     & -0.31 ( 1)  &      --      &      --      & -0.39 ( 8)   & -0.12 ( 2)   & -0.15 ( 3)  & -0.23 ( 8)  & -0.19 ( 1)   & -0.20 ( 1)   \\ 
\rowcolor{lightgray} 																
      &  $\pm$ 0.02  &  $\pm$ 0.02  &  $\pm$ 0.10 &  $\pm$ 0.03 &  $\pm$ 0.02  &  $\pm$ 0.04  &  $\pm$ 0.02 &  $\pm$ 0.10 &  $\pm$ 0.10 &  $\pm$ 0.08 &  $\pm$ 0.02  &  $\pm$ 0.11 &  $\pm$ 0.02  &  $\pm$ 0.10  &      --     &  $\pm$ 0.10 &      --      &      --      &  $\pm$ 0.05  &  $\pm$ 0.10  &  $\pm$ 0.03 &  $\pm$ 0.03 &  $\pm$ 0.10  &  $\pm$ 0.10  \\ 
4     & -0.38 (49)   & -0.35 (21)   & -0.28 ( 3)  & -0.40 ( 3)  & -0.39 (17)   & -0.37 (18)   & -0.35 ( 9)  & -0.26 ( 2)  & -0.38 ( 2)  & -0.42 ( 3)  & -0.34 ( 8)   & -0.28 ( 6)  & -0.39 ( 7)   & -0.24 ( 1)   &      --     & -0.36 ( 2)  & -0.30 ( 1)   & -0.30 ( 1)   & -0.36 ( 7)   & -0.10 ( 2)   & -0.10 ( 3)  & -0.18 ( 8)  & -0.27 ( 1)   & -0.27 ( 1)   \\ 
      &  $\pm$ 0.02  &  $\pm$ 0.02  &  $\pm$ 0.10 &  $\pm$ 0.03 &  $\pm$ 0.01  &  $\pm$ 0.04  &  $\pm$ 0.03 &  $\pm$ 0.10 &  $\pm$ 0.10 &  $\pm$ 0.10 &  $\pm$ 0.02  &  $\pm$ 0.11 &  $\pm$ 0.02  &  $\pm$ 0.10  &      --     &  $\pm$ 0.10 &  $\pm$ 0.10  &  $\pm$ 0.10  &  $\pm$ 0.02  &  $\pm$ 0.10  &  $\pm$ 0.03 &  $\pm$ 0.02 &  $\pm$ 0.10  &  $\pm$ 0.10  \\ 
\rowcolor{lightgray} 																
5     & -0.31 (48)   & -0.36 (21)   & -0.24 ( 3)  & -0.37 ( 3)  & -0.29 (17)   & -0.30 (16)   & -0.29 ( 7)  & -0.34 ( 1)  & -0.39 ( 1)  & -0.26 ( 4)  & -0.21 ( 6)   & -0.26 ( 5)  & -0.32 ( 8)   & -0.24 ( 1)   &      --     & -0.22 ( 2)  & -0.31 ( 1)   & -0.31 ( 1)   & -0.40 ( 7)   & -0.09 ( 2)   & -0.15 ( 3)  & -0.25 ( 6)  & -0.24 ( 1)   &	  --	   \\ 
\rowcolor{lightgray} 																
      &  $\pm$ 0.02  &  $\pm$ 0.02  &  $\pm$ 0.10 &  $\pm$ 0.03 &  $\pm$ 0.01  &  $\pm$ 0.04  &  $\pm$ 0.03 &  $\pm$ 0.10 &  $\pm$ 0.10 &  $\pm$ 0.09 &  $\pm$ 0.02  &  $\pm$ 0.12 &  $\pm$ 0.02  &  $\pm$ 0.10  &      --     &  $\pm$ 0.10 &  $\pm$ 0.10  &  $\pm$ 0.10  &  $\pm$ 0.02  &  $\pm$ 0.10  &  $\pm$ 0.03 &  $\pm$ 0.03 &  $\pm$ 0.10  &	  --	   \\ 
6     & -0.33 (57)   & -0.29 (63)   & -0.34 ( 3)  & -0.33 ( 3)  & -0.36 (17)   & -0.29 (16)   & -0.33 ( 7)  & -0.42 ( 1)  & -0.38 ( 1)  & -0.39 ( 5)  & -0.39 ( 8)   & -0.31 ( 4)  & -0.35 ( 8)   & -0.34 ( 1)   & -0.30 ( 2)  & -0.17 ( 1)  &      --      &      --      & -0.23 ( 8)   & -0.05 ( 2)   & -0.15 ( 2)  & -0.06 ( 4)  & -0.25 ( 1)   &	  --	   \\ 
      &  $\pm$ 0.02  &  $\pm$ 0.01  &  $\pm$ 0.17 &  $\pm$ 0.03 &  $\pm$ 0.02  &  $\pm$ 0.03  &  $\pm$ 0.04 &  $\pm$ 0.10 &  $\pm$ 0.10 &  $\pm$ 0.09 &  $\pm$ 0.05  &  $\pm$ 0.12 &  $\pm$ 0.06  &  $\pm$ 0.10  &  $\pm$ 0.10 &  $\pm$ 0.10 &      --      &      --      &  $\pm$ 0.03  &  $\pm$ 0.10  &  $\pm$ 0.10 &  $\pm$ 0.02 &  $\pm$ 0.10  &	  --	   \\ 
\rowcolor{lightgray} 																 
7     & -0.33 (57)   & -0.29 (63)   & -0.34 ( 3)  & -0.30 ( 2)  & -0.36 (16)   & -0.20 (18)   & -0.33 ( 7)  & -0.38 ( 1)  & -0.38 ( 1)  & -0.25 ( 6)  & -0.39 ( 7)   & -0.29 ( 6)  & -0.35 ( 7)   & -0.20 ( 1)   & -0.25 ( 2)  & -0.15 ( 2)  &      --      &      --      & -0.23 ( 8)   & -0.02 ( 1)   & -0.01 ( 3)  &  0.01 ( 5)  & -0.22 ( 1)   &	  --	   \\ 
\rowcolor{lightgray} 																
      &  $\pm$ 0.02  &  $\pm$ 0.01  &  $\pm$ 0.03 &  $\pm$ 0.10 &  $\pm$ 0.01  &  $\pm$ 0.02  &  $\pm$ 0.03 &  $\pm$ 0.10 &  $\pm$ 0.10 &  $\pm$ 0.06 &  $\pm$ 0.02  &  $\pm$ 0.05 &  $\pm$ 0.03  &  $\pm$ 0.10  &  $\pm$ 0.10 &  $\pm$ 0.10 &      --      &      --      &  $\pm$ 0.03  &  $\pm$ 0.10  &  $\pm$ 0.02 &  $\pm$ 0.02 &  $\pm$ 0.10  &	  --	   \\ 
8     & -0.30 (57)   & -0.30 (21)   & -0.29 ( 3)  & -0.35 ( 1)  & -0.42 (15)   & -0.36 (17)   & -0.39 ( 8)  & -0.19 ( 2)  &      --     & -0.39 ( 2)  & -0.37 ( 8)   & -0.23 ( 3)  & -0.39 ( 5)   &      --      &      --     & -0.31 ( 1)  & -0.30 ( 1)   & -0.30 ( 1)   & -0.31 (10)   & -0.32 ( 1)   & -0.21 ( 3)  & -0.22 (10)  & -0.19 ( 1)   & -0.16 ( 1)   \\ 
      &  $\pm$ 0.02  &  $\pm$ 0.01  &  $\pm$ 0.02 &  $\pm$ 0.10 &  $\pm$ 0.03  &  $\pm$ 0.04  &  $\pm$ 0.03 &  $\pm$ 0.10 &      --     &  $\pm$ 0.10 &  $\pm$ 0.03  &  $\pm$ 0.12 &  $\pm$ 0.03  &      --      &      --     &  $\pm$ 0.10 &  $\pm$ 0.10  &  $\pm$ 0.10  &  $\pm$ 0.02  &  $\pm$ 0.10  &  $\pm$ 0.03 &  $\pm$ 0.03 &  $\pm$ 0.10  &  $\pm$ 0.10  \\ 
\rowcolor{lightgray} 																
9     & -0.29 (56)   & -0.33 (21)   & -0.26 ( 3)  & -0.42 ( 1)  & -0.40 (13)   & -0.37 (17)   & -0.35 ( 7)  & -0.19 ( 2)  &      --     & -0.37 ( 2)  & -0.38 ( 8)   & -0.27 ( 3)  & -0.34 ( 7)   &      --      &      --     & -0.26 ( 2)  & -0.37 ( 1)   & -0.37 ( 1)   & -0.28 (10)   & -0.20 ( 2)   & -0.20 ( 3)  & -0.27 (10)  & -0.23 ( 1)   &	  --	   \\
\rowcolor{lightgray} 																
      &  $\pm$ 0.02  &  $\pm$ 0.02  &  $\pm$ 0.02 &  $\pm$ 0.10 &  $\pm$ 0.03  &  $\pm$ 0.05  &  $\pm$ 0.03 &  $\pm$ 0.10 &      --     &  $\pm$ 0.10 &  $\pm$ 0.02  &  $\pm$ 0.10 &  $\pm$ 0.05  &      --      &      --     &  $\pm$ 0.10 &  $\pm$ 0.10  &  $\pm$ 0.10  &  $\pm$ 0.04  &  $\pm$ 0.10  &  $\pm$ 0.03 &  $\pm$ 0.04 &  $\pm$ 0.10  &	  --	   \\ 
10    & -0.24 (60)   & -0.24 (22)   & -0.25 ( 3)  & -0.27 ( 1)  & -0.24 (16)   & -0.23 (18)   & -0.31 ( 7)  & -0.19 ( 2)  & -0.33 ( 2)  & -0.31 ( 2)  & -0.31 ( 8)   & -0.25 ( 2)  & -0.30 ( 7)   &      --      &      --     & -0.17 ( 1)  & -0.29 ( 1)   & -0.29 ( 1)   & -0.14 (10)   &  0.04 ( 1)   & -0.06 ( 2)  & -0.13 ( 9)  & -0.19 ( 1)   &	  --	   \\ 
      &  $\pm$ 0.02  &  $\pm$ 0.01  &  $\pm$ 0.02 &  $\pm$ 0.10 &  $\pm$ 0.02  &  $\pm$ 0.04  &  $\pm$ 0.04 &  $\pm$ 0.10 &  $\pm$ 0.10 &  $\pm$ 0.10 &  $\pm$ 0.04  &  $\pm$ 0.10 &  $\pm$ 0.02  &      --      &      --     &  $\pm$ 0.10 &  $\pm$ 0.10  &  $\pm$ 0.10  &  $\pm$ 0.04  &  $\pm$ 0.10  &  $\pm$ 0.09 &  $\pm$ 0.03 &  $\pm$ 0.10  &	  --	   \\ 
\rowcolor{lightgray} 																
11    & -0.30 (61)   & -0.31 (20)   & -0.32 ( 2)  & -0.26 ( 1)  & -0.28 (16)   & -0.27 (18)   & -0.28 ( 8)  & -0.27 ( 3)  & -0.41 ( 2)  & -0.34 ( 2)  & -0.32 ( 8)   & -0.19 ( 3)  & -0.38 ( 2)   &      --      &      --     &      --     & -0.38 ( 1)   & -0.38 ( 1)   & -0.24 (10)   & -0.18 ( 1)   & -0.24 ( 3)  & -0.23 ( 9)  & -0.10 ( 1)   &	  --	   \\ 
\rowcolor{lightgray} 																
      &  $\pm$ 0.02  &  $\pm$ 0.01  &  $\pm$ 0.10 &  $\pm$ 0.10 &  $\pm$ 0.02  &  $\pm$ 0.05  &  $\pm$ 0.05 &  $\pm$ 0.16 &  $\pm$ 0.10 &  $\pm$ 0.10 &  $\pm$ 0.02  &  $\pm$ 0.13 &  $\pm$ 0.10  &      --      &      --     &  --         &  $\pm$ 0.10  &  $\pm$ 0.10  &  $\pm$ 0.4  &  $\pm$ 0.10   &  $\pm$ 0.03 &  $\pm$ 0.07 &  $\pm$ 0.10  &	  --	   \\ 
12    & -0.26 (60)   & -0.25 (23)   & -0.29 ( 3)  & -0.25 ( 1)  & -0.28 (15)   & -0.28 (17)   & -0.31 ( 9)  & -0.11 ( 2)  &      --     & -0.31 ( 2)  & -0.27 ( 8)   & -0.32 ( 2)  & -0.34 ( 7)   &      --      &      --     &      --     & -0.22 ( 1)   & -0.22 ( 1)   & -0.17 (10)   &  0.08 ( 2)   &  0.01 ( 1)  & -0.14 (10)  & -0.10 ( 1)   &	  --	   \\ 
      &  $\pm$ 0.02  &  $\pm$ 0.01  &  $\pm$ 0.01 &  $\pm$ 0.10 &  $\pm$ 0.03  &  $\pm$ 0.05  &  $\pm$ 0.02 &  $\pm$ 0.10 &      --     &  $\pm$ 0.10 &  $\pm$ 0.03  &  $\pm$ 0.10 &  $\pm$ 0.5  &      --       &      --     &  --         &  $\pm$ 0.10  &  $\pm$ 0.10  &  $\pm$ 0.04  &  $\pm$ 0.10  &  $\pm$ 0.10 &  $\pm$ 0.04 &  $\pm$ 0.10  &	  --	   \\ 
\rowcolor{lightgray} 																
13    & -0.26 (56)   & -0.25 (23)   & -0.30 ( 3)  & -0.47 ( 1)  & -0.37 ( 9)   & -0.26 (17)   & -0.31 ( 8)  & -0.17 ( 3)  & -0.28 ( 1)  & -0.34 ( 2)  & -0.31 ( 7)   & -0.35 ( 3)  & -0.37 ( 8)   &      --      &      --     &      --     & -0.34 ( 1)   & -0.34 ( 1)   & -0.22 (10)   &  -0.37 ( 2)  & -0.07 ( 1)  & -0.17 ( 9)  & -0.20 ( 1)   &	  --	   \\ 
\rowcolor{lightgray} 																
      &  $\pm$ 0.02  &  $\pm$ 0.01  &  $\pm$ 0.02 &  $\pm$ 0.10 &  $\pm$ 0.03  &  $\pm$ 0.04  &  $\pm$ 0.02 &  $\pm$ 0.14 &  $\pm$ 0.10 &  $\pm$ 0.10 &  $\pm$ 0.03  &  $\pm$ 0.09 &  $\pm$ 0.03  &      --      &      --     &  --         &  $\pm$ 0.10  &  $\pm$ 0.10  &  $\pm$ 0.03  &  $\pm$ 0.10  &  $\pm$ 0.10 &  $\pm$ 0.02 &  $\pm$ 0.10  &	  --	   \\ 
14    & -0.31 (62)   & -0.32 (22)   & -0.33 ( 3)  & -0.25 ( 1)  & -0.28 (12)   & -0.33 (18)   & -0.34 ( 7)  & -0.25 ( 3)  & -0.40 ( 2)  & -0.39 ( 3)  & -0.35 ( 8)   & -0.30 ( 3)  & -0.39 ( 5)   &      --      &      --     & -0.25 ( 2)  & -0.45 ( 1)   & -0.45 ( 1)   & -0.35 (10)   & -0.16 ( 2)   & -0.13 ( 1)  & -0.29 (10)  & -0.17 ( 1)   &	  --	   \\ 
      &  $\pm$ 0.02  &  $\pm$ 0.01  &  $\pm$ 0.03 &  $\pm$ 0.10 &  $\pm$ 0.04  &  $\pm$ 0.04  &  $\pm$ 0.02 &  $\pm$ 0.09 &  $\pm$ 0.10 &  $\pm$ 0.07 &  $\pm$ 0.04  &  $\pm$ 0.14 &  $\pm$ 0.02  &      --      &      --     &  $\pm$ 0.10 &  $\pm$ 0.10  &  $\pm$ 0.10  &  $\pm$ 0.01  &  $\pm$ 0.10  &  $\pm$ 0.10 &  $\pm$ 0.04 &  $\pm$ 0.10  &	  --	   \\ 
\rowcolor{lightgray} 																
15    & -0.28 (54)   & -0.27 (23)   & -0.30 ( 3)  &      --     & -0.27 (10)   & -0.22 (15)   & -0.30 ( 7)  & -0.24 ( 2)  & -0.27 ( 2)  & -0.43 ( 3)  & -0.28 ( 6)   & -0.36 ( 3)  & -0.35 ( 7)   &      --      &      --     &      --     & -0.17 ( 1)   & -0.17 ( 1)   & -0.08 (10)   &  0.17 ( 1)   &	--         & -0.02 ( 8)  & -0.02 ( 1)   &	  --	   \\ 
\rowcolor{lightgray} 																
      &  $\pm$ 0.02  &  $\pm$ 0.02  &  $\pm$ 0.02 &      --     &  $\pm$ 0.04  &  $\pm$ 0.04  &  $\pm$ 0.03 &  $\pm$ 0.10 &  $\pm$ 0.10 &  $\pm$ 0.11 &  $\pm$ 0.05  &  $\pm$ 0.09 &  $\pm$ 0.04  &      --      &      --     &  --         &  $\pm$ 0.10  &  $\pm$ 0.10  &  $\pm$ 0.06  &  $\pm$ 0.10  &	--         &  $\pm$ 0.03 &  $\pm$ 0.10  &	  --	   \\ 
16    & -0.26 (55)   & -0.23 ( 6)   & -0.26 ( 2)  & -0.35 ( 1)  &      --	   & -0.31 (18)   & -0.38 ( 1)  & -0.23 ( 2)  &      --     & -0.33 ( 3)  &      --      & -0.19 ( 3)  &  	   --     &      --      &      --     & -0.32 ( 2)  &      --      &      -       &	  --	  & -0.25 ( 2)   & -0.28 ( 1)  & -0.32 ( 2)  & -0.30 ( 1)   &	  --	   \\ 
      &  $\pm$ 0.02  &  $\pm$ 0.01  &  $\pm$ 0.10 &  $\pm$ 0.10 &      --      &  $\pm$ 0.04  &  $\pm$ 0.10 &  $\pm$ 0.10 &      --     &  $\pm$ 0.04 &      --      &  $\pm$ 0.02 &      --      &      --      &      --     &  $\pm$ 0.10 &  $\pm$ 0.10  &      --      &  	  --	  &  $\pm$ 0.10  &  $\pm$ 0.10 &  $\pm$ 0.10 &  $\pm$ 0.10  &	  --	   \\ 
\rowcolor{lightgray} 																
17    & -0.30 (61)   & -0.32 (19)   & -0.29 ( 3)  & -0.34 ( 1)  & -0.40 (13)   & -0.37 (18)   & -0.36 ( 7)  & -0.24 ( 1)  &      --     & -0.37 ( 3)  & -0.39 ( 7)   & -0.31 ( 3)  & -0.34 ( 6)   &   -0.19 ( 1) &      --     & -0.34 ( 2)  & -0.38 ( 1)   & -0.38 ( 1)   & -0.41 ( 8)   & -0.11 ( 2)   & -0.18 ( 2)  & -0.33 ( 8)  & -0.17 ( 1)   &	  --	   \\ 
\rowcolor{lightgray} 																
      &  $\pm$ 0.02  &  $\pm$ 0.01  &  $\pm$ 0.02 &  $\pm$ 0.10 &  $\pm$ 0.02  &  $\pm$ 0.04  &  $\pm$ 0.02 &  $\pm$ 0.10 &      --     &  $\pm$ 0.10 &  $\pm$ 0.05  &  $\pm$ 0.08 &  $\pm$ 0.02  &  $\pm$ 0.10  &      --     &  $\pm$ 0.10 &  $\pm$ 0.10  &  $\pm$ 0.10  &  $\pm$ 0.03  &  $\pm$ 0.10  &  $\pm$ 0.10 &  $\pm$ 0.05 &  $\pm$ 0.10  &	  --	   \\ 
18    & -0.29 (66)   & -0.26 (52)   & -0.22 ( 3)  & -0.27 ( 1)  & -0.30 (12)   & -0.33 (16)   & -0.33 ( 8)  & -0.16 ( 3)  & -0.08 ( 2)  & -0.48 ( 3)  & -0.34 ( 3)   & -0.29 ( 2)  & -0.34 ( 7)   &   -0.27 ( 1) &      --     &      --     & -0.14 ( 1)   & -0.14 ( 1)   & -0.19 ( 9)   &  0.08 ( 1)   &  0.01 ( 3)  & -0.02 ( 9)  & -0.25 ( 1)   &	  --	   \\ 
      &  $\pm$ 0.02  &  $\pm$ 0.01  &  $\pm$ 0.03 &  $\pm$ 0.10 &  $\pm$ 0.04  &  $\pm$ 0.03  &  $\pm$ 0.03 &  $\pm$ 0.16 &  $\pm$ 0.10 &  $\pm$ 0.10 &  $\pm$ 0.03  &  $\pm$ 0.10 &  $\pm$ 0.02  &  $\pm$ 0.10  &      --     &  --         &  $\pm$ 0.10  &  $\pm$ 0.10  &  $\pm$ 0.04  &  $\pm$ 0.10  &  $\pm$ 0.07 &  $\pm$ 0.03 &  $\pm$ 0.10  &	  --	   \\ 
\rowcolor{lightgray} 																
19    & -0.29 (55)   & -0.27 ( 9)   & -0.30 ( 2)  & -0.40 ( 1)  &      --	   & -0.39 (11)   & -0.42 ( 1)  & -0.15 ( 2)  &      --     & -0.35 ( 3)  &      --      & -0.36 ( 3)  &	   --     &  -0.23 ( 1)  &      --     & -0.31 ( 1)  &      --      &      --      &	  --	  &	 -0.23 ( 2)	 & -0.25 ( 1)  & -0.38 ( 2)  & -0.29 ( 1)   &	  --	   \\ 
\rowcolor{lightgray} 																
      &  $\pm$ 0.02  &  $\pm$ 0.02  &  $\pm$ 0.10 &  $\pm$ 0.10 &      --      &  $\pm$ 0.01  &  $\pm$ 0.10 &  $\pm$ 0.10 &      --     &  $\pm$ 0.15 &      --      &  $\pm$ 0.06 &      --      &  $\pm$ 0.10  &      --     &  $\pm$ 0.10 &      --      &      --      &	  --	  &	 --	         &  $\pm$ 0.10 &  $\pm$ 0.10 &  $\pm$ 0.10  &	  --	   \\ 
20    & -0.25 (59)   & -0.26 (20)   & -0.26 ( 3)  & -0.24 ( 1)  & -0.25 (12)   & -0.25 (17)   & -0.26 ( 7)  & -0.04 ( 2)  & -0.15 ( 1)  & -0.34 ( 2)  & -0.28 ( 8)   & -0.31 ( 2)  & -0.30 ( 7)   &      --      &      --     &      --     & -0.27 ( 1)   & -0.27 ( 1)   & -0.23 ( 4)   &  -0.18 ( 1)  &	--         & -0.16 ( 8)  & -0.15 ( 1)   &	  --	   \\ 
      &  $\pm$ 0.02  &  $\pm$ 0.01  &  $\pm$ 0.02 &  $\pm$ 0.10 &  $\pm$ 0.03  &  $\pm$ 0.04  &  $\pm$ 0.03 &  $\pm$ 0.10 &  $\pm$ 0.10 &  $\pm$ 0.10 &  $\pm$ 0.03  &  $\pm$ 0.10 &  $\pm$ 0.02  &      --      &      --     &  --         &  $\pm$ 0.10  &  $\pm$ 0.10  &  $\pm$ 0.01  &  $\pm$ 0.10  &	--         &  $\pm$ 0.04 &  $\pm$ 0.10  &	  --	   \\ 
\rowcolor{lightgray} 																
21    & -0.29 (61)   & -0.26 ( 6)   & -0.26 ( 2)  & -0.32 ( 1)  &      --	   & -0.37 (17)   & -0.38 ( 1)  & -0.05 ( 2)  &      --     & -0.37 ( 3)  &      --      & -0.23 ( 3)  &	   --     &  -0.16 ( 1)  &      --     &      --     &      --      &      --      &	  --	  &	 --	         & -0.19 ( 1)  & -0.22 ( 2)  & -0.28 ( 1)   &	  --	   \\ 
\rowcolor{lightgray} 																
      &  $\pm$ 0.02  &  $\pm$ 0.01  &  $\pm$ 0.10 &  $\pm$ 0.10 &      --      &  $\pm$ 0.04  &  $\pm$ 0.10 &  $\pm$ 0.10 &      --     &  $\pm$ 0.03 &      --      &  $\pm$ 0.08 &      --      &  $\pm$ 0.10  &      --     &  --         &      --      &      --      &	  --	  &	 --	         &  $\pm$ 0.10 &  $\pm$ 0.10 &  $\pm$ 0.10  &	  --	   \\ 
22    & -0.30 (53)   & -0.34 (18)   & -0.32 ( 2)  & -0.26 ( 1)  &  -0.37 ( 5)  & -0.29 (17)   & -0.36 ( 5)  & -0.31 ( 2)  & -0.33 ( 2)  & -0.38 ( 2)  & -0.35 ( 8)   & -0.18 ( 3)  & -0.29 ( 2)   &   -0.16 ( 1) &      --     &      --     & -0.32 ( 1)   & -0.32 ( 1)   & -0.25 ( 6)   &  -0.10 ( 1)  &	--         & -0.32 ( 6)  & -0.16 ( 1)   &	  --	   \\ 
      &  $\pm$ 0.02  &  $\pm$ 0.03  &  $\pm$ 0.10 &  $\pm$ 0.10 &  $\pm$ 0.03  &  $\pm$ 0.04  &  $\pm$ 0.02 &  $\pm$ 0.10 &  $\pm$ 0.10 &  $\pm$ 0.10 &  $\pm$ 0.03  &  $\pm$ 0.16 &  $\pm$ 0.06  &  $\pm$ 0.10  &      --     &  --         &  $\pm$ 0.10  &  $\pm$ 0.10  &  $\pm$ 0.03  &  $\pm$ 0.10  &	--         &  $\pm$ 0.03 &  $\pm$ 0.10  &	  --	   \\ 
\rowcolor{lightgray} 																
23    & -0.30 (56)   & -0.28 ( 6)   & -0.24 ( 2)  & -0.32 ( 1)  &      --	   & -0.28 (17)   & -0.41 ( 1)  &      --     &      --     & -0.37 ( 3)  &      --      & -0.25 ( 3)  &	   --     &   -0.26 ( 1) &      --     &      --     &      --      &      --      &	  --	  &	 -0.02 ( 1)	 & -0.30 ( 1)  & -0.42 ( 2)  & -0.30 ( 1)   &	  --	   \\ 
\rowcolor{lightgray} 																
      &  $\pm$ 0.02  &  $\pm$ 0.03  &  $\pm$ 0.10 &  $\pm$ 0.10 &      --      &  $\pm$ 0.04  &  $\pm$ 0.10 &  $\pm$ 0.10 &      --     &  $\pm$ 0.06 &      --      &  $\pm$ 0.03 &      --      &  $\pm$ 0.10  &      --     &  --         &      --      &      --      &	  --	  & $\pm$ 0.10   &  $\pm$ 0.10 &  $\pm$ 0.10 &  $\pm$ 0.10  &	  --	   \\ 
24    & -0.28 (54)   & -0.26 (54)   & -0.31 ( 3)  & -0.50 ( 1)  & -0.33 ( 5)   & -0.28 (16)   & -0.37 ( 7)  &  0.00 ( 2)  & -0.15 ( 2)  & -0.42 ( 3)  & -0.33 ( 5)   & -0.29 ( 2)  & -0.29 ( 7)   &   -0.23 ( 1) &      --     &      --     & -0.21 ( 1)   & -0.21 ( 1)   & -0.24 ( 8)   &	 0.08 (1)	 & -0.08 ( 3)  & -0.05 ( 8)  & -0.17 ( 1)   &	  --	   \\ 
      &  $\pm$ 0.02  &  $\pm$ 0.01  &  $\pm$ 0.05 &  $\pm$ 0.10 &  $\pm$ 0.05  &  $\pm$ 0.04  &  $\pm$ 0.02 &  $\pm$ 0.10 &  $\pm$ 0.10 &  $\pm$ 0.10 &  $\pm$ 0.06  &  $\pm$ 0.10 &  $\pm$ 0.04  &  $\pm$ 0.10  &      --     &  --         &  $\pm$ 0.10  &  $\pm$ 0.10  &  $\pm$ 0.03  & $\pm$ 0.10   &  $\pm$ 0.04 &  $\pm$ 0.03 &  $\pm$ 0.10  &	  --	   \\ 
\rowcolor{lightgray} 																
25    & -0.31 (46)   & -0.28 ( 6)   & -0.27 ( 1)  & -0.29 ( 1)  &      --	   & -0.29 (11)   & -0.37 ( 1)  & -0.15 ( 2)  &      --     & -0.25 ( 3)  &      --      & -0.27 ( 2)  &	   --     &  -0.24 ( 1)  &      --     & -0.16 ( 1)  &      --      &      --      &	  --	  &	 0.00 ( 1)	 & -0.22 ( 1)  & -0.31 ( 2)  &      --      &	  --	   \\ 
\rowcolor{lightgray} 																
      &  $\pm$ 0.02  &  $\pm$ 0.02  &  $\pm$ 0.10 &  $\pm$ 0.10 &      --      &  $\pm$ 0.02  &  $\pm$ 0.10 &  $\pm$ 0.10 &      --     &  $\pm$ 0.03 &      --      &  $\pm$ 0.10 &      --      &  $\pm$ 0.10  &      --     &  $\pm$ 0.10 &      --      &      --      &	  --	  &	 $\pm$ 0.10  &  $\pm$ 0.10 &  $\pm$ 0.10 &      --      &	  --	   \\ 
26    & -0.31 (54)   & -0.28 (16)   & -0.34 ( 3)  & -0.34 ( 1)  & -0.38 ( 8)   & -0.37 (14)   & -0.37 ( 7)  & -0.06 ( 2)  &      --     & -0.38 ( 3)  & -0.35 ( 7)   & -0.36 ( 3)  & -0.41 ( 6)   &   -0.13 ( 1) &      --     & -0.36 ( 2)  & -0.26 ( 1)   & -0.26 ( 1)   & -0.35 ( 8)   & -0.04 ( 2)   & -0.14 ( 3)  & -0.33 ( 8)  & -0.24 ( 1)   &	  --	   \\ 
      &  $\pm$ 0.02  &  $\pm$ 0.04  &  $\pm$ 0.02 &  $\pm$ 0.10 &  $\pm$ 0.04  &  $\pm$ 0.04  &  $\pm$ 0.02 &  $\pm$ 0.10 &      --     &  $\pm$ 0.02 &  $\pm$ 0.02  &  $\pm$ 0.10 &  $\pm$ 0.04  &  $\pm$ 0.10  &    --       &  $\pm$ 0.10 &  $\pm$ 0.10  &  $\pm$ 0.10  &  $\pm$ 0.03  &  $\pm$ 0.10  &  $\pm$ 0.03 &  $\pm$ 0.03 &  $\pm$ 0.10  &	  --	   \\ 
\rowcolor{lightgray} 																
27    & -0.28 (36)   & -0.26 ( 6)   & -0.24 ( 2)  & -0.36 ( 1)  &      --	   & -0.34 (13)   & -0.38 ( 1)  & -0.18 ( 2)  &      --     & -0.33 ( 3)  &      --      & -0.30 ( 3)  &	   --     &   -0.23 ( 1) &      --     & -0.20 ( 2)  &      --      &      --      &	  --	  & -0.08 ( 1)   & -0.17 ( 1)  & -0.11 ( 2)  &      --      &	  --	   \\ 
\rowcolor{lightgray} 																
      &  $\pm$ 0.01  &  $\pm$ 0.03  &  $\pm$ 0.10 &  $\pm$ 0.10 &      --      &  $\pm$ 0.02  &  $\pm$ 0.10 &  $\pm$ 0.10 &      --     &  $\pm$ 0.03 &      --      &  $\pm$ 0.03 &      --      &  $\pm$ 0.10  &      --     &  $\pm$ 0.10 &      --      &      --      &	  --	  &  $\pm$ 0.10  &  $\pm$ 0.10 &  $\pm$ 0.10 &      --      &	  --	   \\ 
\hline\hline
\end{tabular}
\label{tab_elem2}
\end{table}
\end{landscape}

Measurement errors quoted in Table~\ref{tab_elem1} and Table~\ref{tab_elem2} include uncertainty in the continuum positioning and photon noise and correspond to the $1\sigma$ dispersion divided by the square root of the number of used lines. 
For elements with more than two measurable lines, we computed the dispersion around the mean abundance, while for those with one or two measurable lines, we assumed an error of 0.1~dex.
Abundance errors from uncertainties in the stellar parameters (see Sect.~\ref{stel_param}) were estimated by computing elemental abundances with varying $T_{eff}$ by $\pm100$~K, log(g) by $\pm0.15$~dex, and $\xi$ by $\pm0.2$~$km s^{-1}$. 

Typically, the impact of these uncertainties in the estimated abundances is within 0.10-0.15 dex for all the species with the exception of nitrogen from CN lines, for which it can raise to 0.2 dex. \\
In the computation of these errors we did not include the interdependence between the C, N, and O abundances that contribute to the formation of the measured molecular lines.
We estimate that this effect yields abundance errors below 0.1 dex for C and O, while for N errors can be as high as 0.2-0.3 dex \citep[see also][]{Ryde09}.
We also neglect the effect of the abundances of the main electron donors on those derived from ionised species. However, we estimate that this effect yields abundance errors below 0.1 dex for all the species.
\begin{figure}
    \includegraphics[scale=0.44]{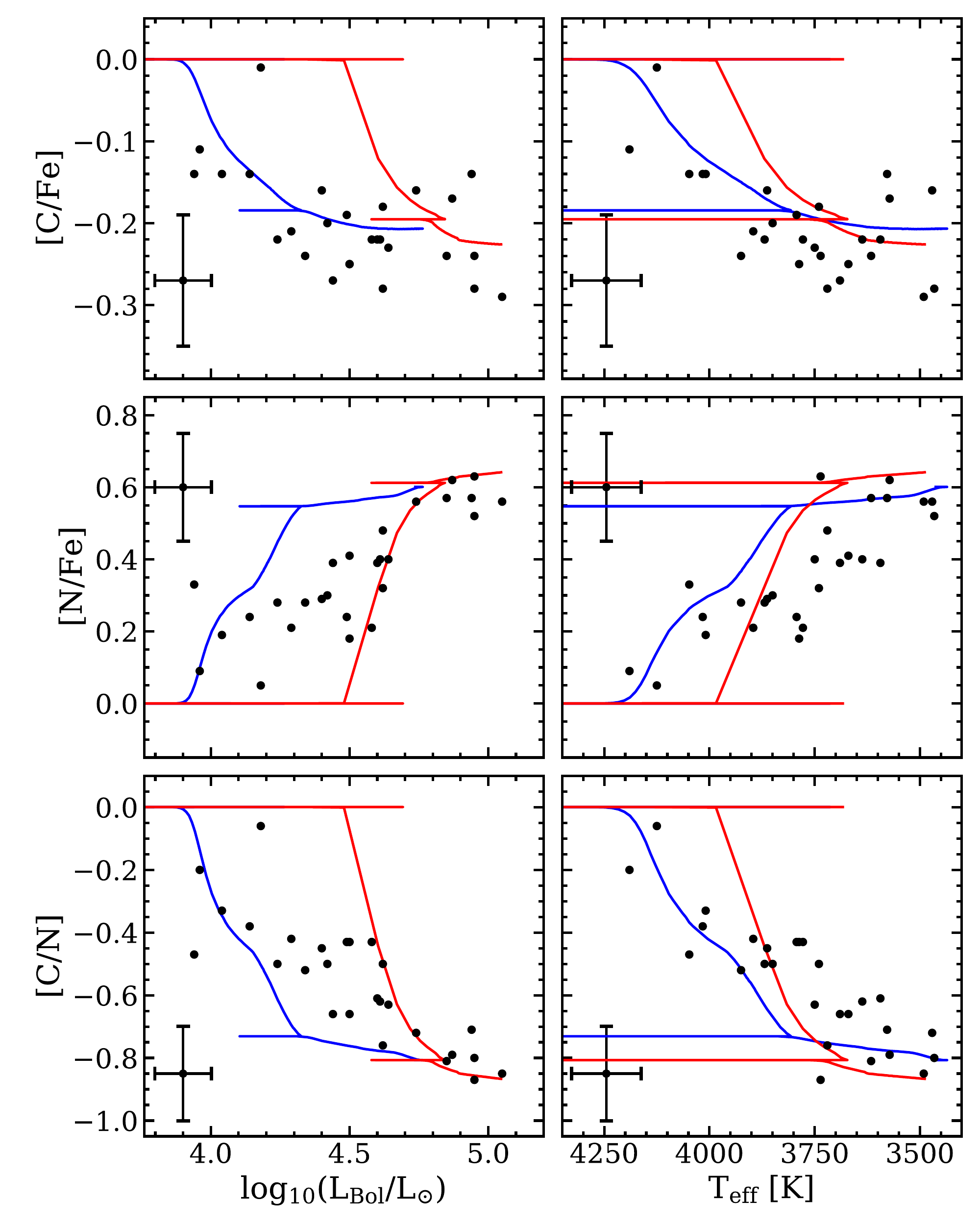}
    \caption{[C/Fe], [N/Fe], [C/N] abundance ratios as a function of the stellar luminosity (left panels) and temperature (right panels) for the 27 studied RSGs in the Perseus complex. Values are solar-scaled according to \citet{Grevesse98} solar reference. 
    Typical errorbars are reported in the left upper or lower corner of each panel.
    For comparison, the PARSEC model predictions for a 9~M$_{\odot}$ (blue lines) and a 14~M$_{\odot}$ (red lines) star have been also plotted.}
    \label{CNrat}
\end{figure}
\begin{figure*}
    \centering
    \includegraphics[scale=0.50]{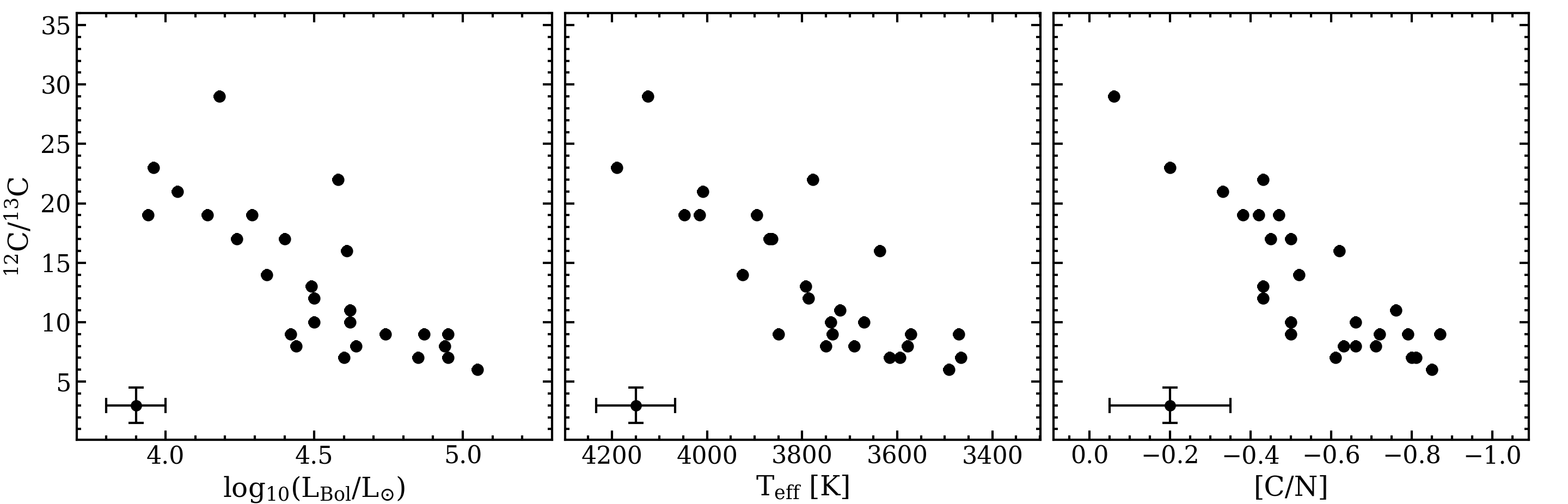}
    \caption{$^{12}$C/$^{13}$C ratio as a function of bolometric luminosity (left panel), effective temperature (central panel) and [C/N] abundance ratio. Typical errorbars are reported in the left lower corner of each panel.}
    \label{c12c13}
\end{figure*}
\section{Discussion}
\label{disc}

The 27 studied RSGs  in the Perseus complex provide an average half-solar iron abundance. This value fits within the metallicity distribution of older dwarf and giant stars \citep[see e.g.][]{Reddy03,hayden14,miko14,miko19} at a similar Galactocentric distance of about 10 kpc, which may imply inflows of metal poor gas as the origin of the sub-solar metallicity in such a young stellar population.
However, the dispersion in metallicity of the measured RSGs in the Perseus complex is significantly lower than the typical values of $0.15-0.20$~dex  measured in the thin disc. Such a  chemical homogeneity reinforces the suggestion by \citealt{dalessandro21}, based on the co-moving kinematics of the LISCA~I stellar sub-structures, that we may have caught an ongoing process of hierarchical cluster assembly.

For the three stars in common with \citet{gazak14} study, we inferred spectroscopic  temperatures that are systematically about 300-400 K cooler than in \citet{gazak14} and in agreement within about 100 K with  those of \citet{Levesque05}, and 0.1-0.3 dex lower metallicities. We also measured some depletion of [C/Fe] and enhancement of [N/Fe] compared to the solar-scaled values adopted by \citet{gazak14}. 
The warm temperatures, about solar metallicity and solar-scaled CNO abundances quoted by \citealt{gazak14} cannot provide a reasonable, simultaneous fit of the CO, OH, CN and TiO molecular absorptions observed in our HARPS-N and GIANO-B spectra of the three RSGs in common.
Indeed, global spectral synthesis in the J-band alone, as used by \citet{gazak14}, may be not sufficiently adequate to simultaneously measure stellar parameters and global metallicity of complex atmospheres like those of  RSGs, since the resulting estimates are likely affected by degeneracy and systematics. In order to mitigate them, a wide NIR  spectral coverage and self-consistent determination of CNO abundances are especially critical requirements.

About solar-scaled [X/Fe] abundance ratios for most of the elements have been measured, consistently with RSGs being a thin disc population. However we also found deviations (on average between 0.1 and 0.2 dex) from solar-scaled abundance ratios for some elements, suggesting possible peculiarities in the recent enrichment of the thin disc by AGB and/or Wolf-Rayet stars within the Perseus complex. 
In particular, an average mild depletion of [F/Fe] and enhancement of [K/Fe] and [Na/Fe] light elements, at variance with the solar-scaled value of [Al/Fe] and [$\alpha$/Fe] elements and a mild enhancement of [Ce/Fe] and [Nd/Fe] heavy-s process elements, at variance with the solar-scaled value of the [Y/Fe] light s-process element have been measured.

Depletion of carbon and of the $^{12}$C/$^{13}$C isotopic abundance ratio and a corresponding enhancement of N with respect to the solar scaled values have been also measured in the 27 studied RSGs, consistent with mixing processes in the stellar interiors, that modify the surface abundances of these elements during the post-MS evolution. Stars with progenitor mass in the 9-14 M$_{\odot}$ range, as for the studied RSGs in the Perseus complex, are indeed expected to have already underwent substantial mixing from the first dredge-up when evolving in the RSG phase.

Fig.~\ref{CNrat} shows the trend of [C/Fe], [N/Fe], [C/N] abundance ratios with varying the bolometric luminosity and the effective temperature.
The rather homogeneous carbon depletion with mild (if any) dependence on the star luminosity and temperature is at variance with the amount of nitrogen enhancement that shows a clear trend with stellar parameters. In particular [N/Fe] increases  with increasing luminosity and decreasing temperature. 
A corresponding larger amount of [C/N] depletion with increasing luminosity and decreasing temperature has been derived, with an average value of [C/N]=-0.56$\pm0.04$ dex  and a dispersion of  $\sigma$=0.20$\pm0.03$ dex.

Different stellar evolution models \citep[see e.g.][]{bertelli09,bres12,georgy13,chieffi13,choi16,davies19} predict C depletion by about 0.2 dex after the first dredge-up, in good agreement with the measured values. Models also predict N enhancement by about 0.5-0.6 dex after the first dredge-up. 
Only the most luminous and coolest RSGs in our sample show such a nitrogen enhancement, while the majority of them show [N/Fe] values increasing from 0.1 to 0.4 dex  with increasing luminosity and decreasing temperature.  
Models also predict a corresponding [C/N] depletion of 0.70-0.80 dex after the first dredge-up, 
as measured only in the most luminous, coolest RSGs of our sample.
This may suggest that N enhancement is more dependent on the stellar mass than the C depletion
\citep[see also][]{davies19}.

As shown in Fig.~\ref{c12c13}, we also measured the $^{12}$C/$^{13}$C isotopic ratio in all the 27 RSGs analyzed, finding values that progressively decreases from a few tens to below ten with increasing luminosity and decreasing temperature. 
Only the warmest RSGs show isotopic ratios above 15 as predicted by the majority of models after the first dredge-up, while most of the RSGs with T$_{eff}<$3800~K show values below 10, possibly requiring some extra-mixing to be explained. Only the \citet{choi16} models can possibly reach values as low as 5-10.
We also find a nice correlation between depletion of  $^{12}$C/$^{13}$C and of [C/N].

\section{Conclusions}
\label{conclusions}

{For the first time a combined high resolution optical and NIR chemical analysis of RSGs has been provided, finding an excellent agreement between optical and NIR abundances for all the elements in common.}
The present study has shown the effectiveness of RSGs of K and M spectral types in tracing the detailed chemistry and recent enrichment of their host, by providing a comprehensive characterization of the chemistry of the young stellar populations in the Perseus complex in the outer Galaxy disc. 

To this purpose, the availability of high resolution, high signal-to-noise spectra in the Y,J,H and K NIR bands has been  crucial to  derive accurate stellar parameters and chemical abundances for the full set of of iron, iron-peak, CNO, alpha and other light and neutron-capture elements. In particular, abundances for Na, Mg, Al, Si, S, K, Ca, Sc, Ti, V, Cr, Mn, Fe, Co, Ni, Cu, Zn\textbf{,} Y, Ce, Dy from atomic lines and C, N, O and F from molecular lines.

Despite the low temperatures and gravities of RSGs, optical spectra in a few selected windows less affected by molecular blending and blanketing can also provide some lines for chemical analysis of a sub-set of elements, namely O, Na, Mg, Al, Si, Ca, Sc, Ti, V, Cr, Mn, Fe, Co, Ni\textbf{,} Y, Ce, Dy, and for two neutron-capture elements, namely Zr and Nd, not easily derivable from the few available IR lines, that turn out to be either blended in these cool stars or with uncertain atomic parameters.

When coupled with kinematic information (line-of-sight radial velocities and proper motions), such a detailed chemical screening can enable a consistent  chemo-dynamical modeling and a comprehensive description of modes and timescales of star formation, dynamical interactions and chemical enrichment of the region.

This study can also represent an interesting observational test bench to probe models of stellar evolution, nucleosynthesis and internal mixing in the critical RSG evolutionary phase for stars at half-solar metallicity.
\newline\newline

{\it Acknowledgements.} 
We thank the anonymous referee for his/her detailed report and useful comments and suggestions. We acknowledge the support by INAF/Frontiera through the "Progetti Premiali" funding scheme of the Italian Ministry of Education, University, and Research. We acknowledge support from the project Light-on-Dark granted by MIUR through PRIN2017-000000 contract and support from the mainstream project SC3K - Star clusters in the inner 3 kpc funded by INAF.

\bibliographystyle{aa}
\bibliography{mybib}

\end{document}